\documentclass[10pt,journal,compsoc]{IEEEtran}
\usepackage{bm}
\usepackage{amsmath}
\usepackage{graphicx}
\usepackage{braket}
\usepackage{booktabs}
\usepackage{caption}
\usepackage{subfig}
\usepackage{array}
\usepackage{enumitem}
\usepackage{balance}

\ifCLASSOPTIONcompsoc
  \usepackage[nocompress]{cite}
\else
  \usepackage{cite}
\fi
\ifCLASSINFOpdf
\else
\fi
\usepackage{xcolor,color}

\newcommand{\msf}[1]{\mathsf{#1}}

\begin{document}
%
\title{ViFi: Virtual Fingerprinting WiFi-based Indoor Positioning via Multi-Wall Multi-Floor Propagation Model}
%
%
%
%

\author{Giuseppe~Caso,~\IEEEmembership{Member,~IEEE,}
        Luca~De~Nardis,~\IEEEmembership{Member,~IEEE,}
        Filip~Lemic,~\IEEEmembership{Member,~IEEE,}
        Vlado~Handziski,~\IEEEmembership{Member,~IEEE,}
        Adam~Wolisz,~\IEEEmembership{Senior~Member,~IEEE,}
        and~Maria-Gabriella~Di~Benedetto,~\IEEEmembership{Fellow,~IEEE}
\IEEEcompsocitemizethanks{\IEEEcompsocthanksitem Giuseppe Caso is with the Dept. of Mobile Systems and Analytics (MOSAIC), Simula Metropolitan Center for Digital Engineering (SimulaMet), Oslo, Norway. Previously with the Dept. of Information Engineering, Electronics and Telecommunications (DIET), Sapienza University of Rome, Rome, Italy.\protect\\
E-mail: giuseppe@simula.no, giuseppe.caso@uniroma1.it
\IEEEcompsocthanksitem Luca De Nardis and Maria-Gabriella Di Benedetto are with the Dept. of Information Engineering, Electronics and Telecommunications (DIET), Sapienza University of Rome, Rome,
Italy.\protect\\
E-mail:\{luca.denardis, mariagabriella.dibenedetto\}@uniroma1.it
\IEEEcompsocthanksitem Filip Lemic is with the Internet Technology and Data Science Lab (IDLab), University of Antwerp -- imec, Artwerp, Belgium.\protect\\
E-mail: filip.lemic@uantwerpen.be
\IEEEcompsocthanksitem Vlado Handziski and Adam Wolisz are with the Dept. of Telecommunication Systems, Telecommunication Networks Group (TKN), Technical University of Berlin, Berlin, Germany.\protect\\
E-mail: \{handziski, wolisz\}@tkn.tu-berlin.de}
\thanks{Manuscript received --; revised --.}}

%
%

\markboth{Accepted to IEEE Transactions on Mobile Computing, March 2019}%
{Shell \MakeLowercase{\textit{et al.}}: Multi-Wall Multi-Floor Indoor Propagation Modeling for WiFi Virtual Fingerprinting Indoor Positioning}
%



\IEEEtitleabstractindextext{%
\begin{abstract}
Widespread adoption of indoor positioning systems based on WiFi fingerprinting is at present hindered by the large efforts required for measurements collection during the offline phase. Two approaches were recently proposed to address such issue: crowdsourcing and RSS radiomap prediction, based on either interpolation or propagation channel model fitting from a small set of measurements. RSS prediction promises better positioning accuracy when compared to crowdsourcing, but no systematic analysis of the impact of system parameters on positioning accuracy is available.\\
This paper fills this gap by introducing ViFi, an indoor positioning system that relies on RSS prediction based on Multi-Wall Multi-Floor (MWMF) propagation model to generate a discrete RSS radiomap (\emph{virtual} fingerprints). Extensive experimental results, obtained in multiple independent testbeds, show that ViFi outperforms virtual fingerprinting systems adopting simpler propagation models in terms of accuracy, and allows a sevenfold reduction in the number of measurements to be collected, while achieving the same accuracy of a traditional fingerprinting system deployed in the same environment. Finally, a set of guidelines for the implementation of ViFi in a generic environment, that saves the effort of collecting additional measurements for system testing and fine tuning, is proposed.
\end{abstract}

\begin{IEEEkeywords}
Indoor Positioning, WiFi Fingerprinting, Indoor Propagation Modeling, Multi-Wall Multi-Floor Model, Crowdsourcing.
\end{IEEEkeywords}}

\maketitle

\IEEEdisplaynontitleabstractindextext

%
\IEEEpeerreviewmaketitle

\IEEEraisesectionheading{\section{Introduction}
\label{sec:introduction}}

%
%
%
%
\IEEEPARstart{I}{ndoor} positioning and navigation is a market with expected size of USD 4 billions in 2019 \cite{Markets:2016}, and both research community and industry are currently investing huge efforts in the search for a simple and yet reliable solution to determine the position of a mobile device in an indoor environment. Received Signal Strength (RSS) fingerprinting based on WiFi technology emerged as one of the most popular approaches for the implementation of Indoor Positioning Systems (IPSs) \cite{He:2016}, \cite{Honkavirta:2009}, for two main reasons: 1) it allows to leverage existing communications infrastructure; 2) it can support almost any user device by simply installing an application, making it much easier to deploy than other proposals, that may achieve submeter accuracy but require specific hardware/software modifications to devices \cite{Kotaru:2015}, \cite{Vasisht:2016}.\\
WiFi fingerprinting traditionally operates in two phases. During the so-called \emph{offline phase}, RSS values (\emph{fingerprints}) from WiFi Access Points (APs) detected in the environment are collected at selected positions, referred to as Reference Points (RPs), in order to create a discrete RSS \emph{radiomap} of the area of interest. Within the subsequent \emph{online phase}, the location of target devices is estimated as a function of the positions of the RPs, that best match the RSS values measured by the devices. Accuracy and complexity of fingerprinting mainly depend on two issues: 1) proper definition of similarity metrics and estimation algorithms for selecting the best matching RPs to be used during the online phase, and 2) careful planning of the offline phase, particularly in terms of cardinality and positions of the RPs, and number of measurements collected at each RP \cite{Bahl:2000}, \cite{Kessel:2011}.\\
Regarding the online phase, deterministic and probabilistic $k$-Nearest Neighbors ($k$NN) and Weighted $k$NN (W$k$NN) algorithms are by far the most widely proposed and investigated: on the one hand, deterministic algorithms are appealing and relatively easy to implement, because they take advantage of easily computable deterministic similarity metrics \cite{Li:2006}, \cite{Yu:2014}; on the other hand, probabilistic algorithms may improve the deterministic accuracy at the price of higher computational complexity and measurement efforts, due to the need of reliably estimating the RSS probability distributions in each RP from each WiFi AP \cite{Roos:2002}, \cite{Youssef:2005}, \cite{LeDortz:2012}.\\
The offline phase poses however major issues in large scale deployment of WiFi fingerprinting: the collection of measurements is increasingly time consuming with the area covered by the IPS, and requires on site measurement campaigns that potentially interfere with the activities that are usually carried out in the area. WiFi fingerprinting may therefore be difficult to deploy in cases such as:
\begin{description}[style=unboxed, leftmargin = 0cm]
\item[security-sensitive facilities,] characterized by restricted access areas, where measurements cannot be collected;
\item[emergency and diagnostic rooms in hospitals,] where activities cannot be interrupted by measurement collection;
\item[large shopping malls and skyscrapers] spanning over tens or even hundreds of floors, where the sheer size and extension make the collection efforts nonviable due to costs or time considerations.
\end{description}

\noindent Incidentally however, the above environments are some of the most appealing targets for an IPS system \cite{Du:2016}, \cite{Mathisen:2016}, \cite{Yasmine:2016}. As a consequence, methods for either eliminating or reducing the offline phase were proposed.\\ 
As regards the elimination of the offline phase, most of the works in the literature proposed strategies to model a direct relation between RSS values and spatial coordinates, in order to infer the distances between APs and target device, and in turn target position. Simple radio propagation modeling and regression \cite{Gwon:2004}, \cite{Chintalapudi:2010}, along with either modifications of off-the-shelf devices \cite{Gwon:2004}, or additional hardware \cite{Lim:2006}, \cite{Ji:2006}, \cite{Wu:2013}, were proposed. However, the increase of positioning error \cite{Gwon:2004}, \cite{Chintalapudi:2010}, and the need for enhanced system components \cite{Lim:2006}, \cite{Ji:2006}, \cite{Wu:2013}, challenge the adoption of such approaches in place of traditional fingerprinting.\\
Considering the reduction of the offline phase, two main solutions emerged: 
\begin{description}[style=unboxed, leftmargin = 0cm]
\item [Crowdsourcing] -- In this approach, devices that make use of the IPS contribute to the collection of location-dependent RSS samples \cite{Bolliger:2008}, in either a voluntary or involuntary way \cite{Rai:2012}. 
Application of crowdsourcing entails further challenges, in terms of uncertainty of location of RPs and corresponding RSS values, caused by lack of control on the mobile device, and heterogeneity of devices used for collection \cite{Chatzimilioudis:2012}, \cite{Yang:2013}, \cite{Laoudias:2013}.
\item [Virtual Fingerprinting] -- In this approach, the generation of so-called \emph{virtual} RPs\footnote{\emph{real} RPs indicate real fingerprints collected at the corresponding RP locations, while \emph{virtual} RPs indicate virtual fingerprints generated as a function of the selected RSS prediction method and the corresponding RP locations.} is carried out in the area of interest, by adopting RSS prediction methods based on either \emph{propagation modeling}, in which an empirical radio propagation model is trained with an initial set of measurements, or by \emph{interpolation}, in which adjacent \emph{real} RPs are interpolated. In both cases, virtual RPs, along with real RPs, whose amount is hopefully smaller when compared to traditional fingerprinting, are then jointly used in the online phase.
\end{description}
\noindent In this paper, a WiFi IPS based on \emph{virtual fingerprinting by propagation modeling}, referred to as ViFi, is proposed and analyzed. The system adopts RSS propagation modeling for the offline phase reduction, and combines it with a deterministic position estimation algorithm.\\
Design strategies adopted for ViFi were principally driven by the following considerations:
\begin{enumerate}
\item Differently from offline phase elimination schemes discussed above, virtual fingerprinting does not directly infer distances. This in turn allows the use of advanced propagation models, embedded with multiple propagation and topological parameters, that may lead to improved prediction of RSS values, accurate generation of virtual RPs, and thus enhanced positioning accuracy.
\item In contrast to offline phase reduction schemes by either crowdsourcing or virtual fingerprinting by interpolation, virtual fingerprinting by propagation modeling allows the deployment of IPSs in scenarios where an extensive and spatially regular data collection during the offline phase is prohibitive or even impossible. RSS prediction by advanced propagation models can be in fact obtained even in areas where no measurements are collected, if measurements in environments with similar characteristics, e.g. different floors in the same building, are available, as analyzed in \cite{Caso:2017}, \cite{Lemic:2016}.
\end{enumerate} 
Overall, ViFi offline and online phases were designed aiming to limit the complexity and avoid device modifications and additional hardware. In particular, 1) the Multi-Wall Multi-Floor (MWMF) was selected among several empirical models, since it provides advanced propagation modeling at a reasonable degree of complexity \cite{Caso:2017}, \cite{Widyawan:2007}, \cite{Damosso:1999}, \cite{Borrelli:2004}, \cite{Caso:Fab:2015}, and 2) a traditional, low-complex W$k$NN algorithm was adopted for the fingerprinting online phase.\\
ViFi was implemented and tested in two testbeds with different AP topology and signal coverage, in order to obtain consolidated data supporting the proposed approach, and explore the possibility of introducing a set of guidelines for its seamless deployment in other testbeds. The validity of the proposed guidelines was then corroborated by deploying ViFi in a third testbed, with results presented in \cite{ViFi:supplemental}.\\
The paper is organized as follows: Section \ref{related} reviews related work in the field of RSS prediction, and identifies open issues and contributions of this work beyond the current state of the art. Section \ref{sec:MWMF} contains the analytic foundations of the MWMF indoor propagation model, adopted for the generation of virtual RPs. The ViFi system is described in Section \ref{sec:Phases}, where design choices related to both offline and online phases are discussed. The testbeds used for the experimental analysis are presented and compared in Section \ref{testbeds}, with results being discussed in Section \ref{results}. Section \ref{guidelines} proposes the system implementation guidelines. Finally, Section \ref{requirements} compares ViFi with traditional fingerprinting in terms of design requirements, positioning accuracy, complexity, and vulnerability, while Section \ref{conclusion} concludes the paper and highlights possible future research lines.

\section{Virtual fingerprinting by RSS prediction}
\label{related}
\subsection{RSS Prediction by Interpolation}
\label{related_interpolation}
Virtual fingerprinting with Inverse Distance Weighting (IDW) and Universal Kriging (UK) interpolation schemes was proposed in \cite{Li:2006}, while a linear regression approach was analyzed in \cite{Hossain:2007}. Experimental results showed, in both cases, a positioning error decrease when compared with fingerprinting systems using a low amount of real RPs. Moreover, virtual fingerprinting via Support Vector Regression (SVR) and Gaussian Processes (GP) was recently proposed in \cite{Hernandez:2015} and \cite{Kumar:2016}, respectively. In the first case, a slightly better accuracy with respect to a traditional $k$NN estimator was obtained; in the second case, a $30 \%$ accuracy improvement, with respect to the Horus system \cite{Youssef:2005}, was achieved. As already discussed, a spatially regular collection of real RPs is in general required for the application of interpolation techniques.\\ 
\subsection{RSS Prediction by Indoor Propagation Modeling}
\label{related_model}
An empirical propagation model taking into account the effect of obstructing walls on the perceived RSS was preliminary proposed in the seminal work on WiFi fingerprinting \cite{Bahl:2000}. The so-called Wall Attenuation Factor (WAF) was derived by averaging the differences between Line of Sight (LoS) and Non LoS (NLoS) measurements, with a known and variable amount of obstructing walls in the latter case. Once the WAF was evaluated, other propagation model parameters (path loss at a reference distance and path loss exponent) were computed via linear regression, and RSS values were predicted by using the resulting model. Experimental results showed an increase of the median positioning error in the order of $46 \%$ when such model was used in place of real measurements, calling for the definition of a more accurate propagation model. In \cite{Widyawan:2007} the impact of using virtual fingerprinting on the achievable positioning accuracy was comparatively analyzed: two empirical propagation models (log-distance and MWMF) and a semi-deterministic model (Motif) were used in the RSS prediction and virtual fingerprints generation. Positioning accuracy was then tested for deterministic NN and probabilistic Bayesian estimators, showing an increase of about $30 \%$ in terms of average positioning error for the log-distance model, and a decrease of about $10 \%$ for MWMF and Motif ones, with respect to the NN algorithm adopting real measurements. No analysis on the impact of the amount of real and virtual RPs was however provided and, in addition, little detail was given on the set of propagation parameters used in the MWMF model. Recently, the possibility of generating reliable virtual RPs, through empirical fitting of a simple propagation model, was confirmed in \cite{Pulkkinen:2015}, where a log-distance model was adopted, limiting the optimization to the path loss exponent. A more complex model was proposed in \cite{Kumar:2015}, foreseeing the sum of two exponentials, with a total set of four parameters to be estimated, but the model was not used to reduce the number of offline measurements.\\ 
\subsection{Open issues and proposed contribution}
\label{motivation}
The focal issue emerging from the analysis of previous work is related to the difficulty in achieving a satisfactory trade off between complexity and accuracy of fingerprinting systems. On the one hand, as observed in Section \ref{sec:introduction}, the elimination of the offline phase either leads to accuracy decrease or requires enhanced components, while crowdsourcing introduces further challenges. One the other hand, virtual fingerprinting may provide reasonable solutions to significantly reduce the offline phase, but strong experimental evidence is still missing. As a matter of fact, given a propagation model or an interpolation technique, the impact of number and positions of a) real RPs on the RSS prediction accuracy, and b) virtual RPs on the positioning accuracy is still unclear.\\
As a second issue, it can be observed that, in general, no guidelines are provided for setting up a virtual fingerprinting system in different environments, given that further investigations are required in order to uncover the relation between offline and online system parameters. This limitation leads to a case by case choice rather than systematic settings, e.g. in terms of number, position and spatial distribution of collected real RPs, generated virtual RPs, and the value to assign to $k$ in case a $k$NN/W$k$NN estimator is adopted.\\   
A third issue regards the need for sets of measurements acquired in a controlled and reliable fashion, common to all proposals reviewed in Sections \ref{related_interpolation} and \ref{related_model}. This is in contrast with recent proposals that allow for data collection in an uncontrolled fashion, such  as crowdsourcing. The joint use of crowdsourcing with virtual fingerprinting is indeed all but unexplored in the literature, and its impact on RSS prediction and positioning accuracy is unknown.\\  
As regards the first issue, the contribution of this paper is the definition of a novel virtual fingerprinting system, ViFi, that provides an accuracy comparable to real fingerprinting systems based on exaustive RSS collections, while significantly reducing the amount of data to be collected. This is obtained by adopting the MWMF model to generate virtual RPs, that has a complexity lower than the techniques in \cite{Hernandez:2015}, \cite{Kumar:2016}, and comparable to the models proposed in \cite{Bahl:2000}, \cite{Pulkkinen:2015}, while leading to better RSS prediction. With respect to \cite{Li:2006}, \cite{Widyawan:2007}, \cite{Hossain:2007}, this paper proposes an in-depth experimental analysis on both adopted propagation model and proposed system, focusing on the impact of densities and positions of real and virtual RPs on RSS prediction and positioning accuracy.\\
As regards the second issue, the relation between system parameters is analyzed, and an empirical and effective rule for the selection of the value of $k$ to be used in the W$k$NN estimator, adopted in ViFi for the online phase, is derived and experimentally validated. Altogether, thanks to an extensive experimental analysis, this paper proposes a full set of guidelines for the implementation and setup of ViFi in different environments, indicating: a) the required densities of both real and virtual RPs, b) the strategies to select and place such RPs, and c) the value of $k$, guaranteeing satisfactory performance without requiring a training phase.\\
With respect to the last issue, this paper provides preliminary results on the combination of virtual fingerprinting and crowdsourcing techniques for offline phase reduction, focusing in particular on the aspect of uncertainty of RSS values, inherent to crowdsourcing, and thus offering a framework for further research activities.
\section{Multi-Wall Multi-Floor Propagation Models}
\label{sec:MWMF}
Multi-Wall Multi-Floor models \cite{Damosso:1999} emerged among empirical narrow-band models as an appealing solution for indoor propagation modeling,  due to the good trade off they provide between analytic simplicity and path loss modeling accuracy \cite{Borrelli:2004}. MWMF models take into account objects, that obstruct signal propagation in an indoor wireless link, leading to the following path loss model \cite{Damosso:1999}:

\begin{equation}
\label{mwmf_gen}
\msf{PL}_{\msf{MWMF}} = \msf{PL}_{\msf{OS}} + A_{\msf{MWMF}} \quad [\msf{dB}],
\end{equation}

\noindent where the $\msf{PL}_{\msf{OS}}$ term models the path loss over the Tx-Rx distance $d$, while $A_{\msf{MWMF}}$ models the additional loss due to obstructing obstacles. The $\msf{PL}_{\msf{OS}}$ term is defined according to the One Slope (OS) model, as follows \cite{Damosso:1999}:

\begin{equation}
\label{oneslope}
\msf{PL}_{\msf{OS}}(d,\gamma) = l_{0} + 10\gamma\log(d) \quad [\msf{dB}],
\end{equation}    

\noindent with $l_{0}$ modeling the $d=1$ $\msf{m}$ reference path loss, while $\gamma$ is the path loss exponent (for free space conditions, $\gamma = 2$ and $l_{0} \approx 40.22$ $\msf{dB}$ @ 2.45 GHz \cite{Borrelli:2004}). When, in the most general case, Tx and Rx are located on different floors, $A_{\msf{MWMF}}$ is given by \cite{Damosso:1999}: 

\begin{equation}
\label{cost213}
A_{\msf{MWMF}} = l_{c} + \sum_{n=1}^{N_{\msf{obj}}}\sum_{i=1}^{I_{n}}N_{n,i}l_{n,i} + N_{\msf{f}}^{\big[\frac{N_{\msf{f}}+2}{N_{\msf{f}}+1}-b\big]}l_{\msf{f}} \quad [\msf{dB}], 
\end{equation} 

\noindent the parameters of which are described in TABLE \ref{parameters}.

\begin{table}[!ht]
\centering
\caption{$A_{\msf{MWMF}}$ parameters description.}
\resizebox{\columnwidth}{!}{
\begin{tabular}{p{0.15\textwidth}l}
\toprule
\textbf{Parameter} & \textbf{Description} \\
\midrule
$l_{c}$ & Constant Loss \\
\midrule
$N_{\msf{obj}}$ & Number of different families of 2D objects \\
\midrule
$I_{n}$ &  Number of types of 2D objects considered for family $n$ \\
\midrule
$N_{n,i}$ & Number of 2D obstructing objects of family $n$ and type $i$ \\
\midrule
$N_{\msf{f}}$ & Number of obstructing floors \\
\midrule
$l_{n,i}$ & Loss due to 2D objects of family $n$ and type $i$ \\
\midrule
$l_{\msf{f}}$ & Loss due to obstructing floors \\
\midrule
$b$ & Empirical 3D propagation parameter \\
\bottomrule
\end{tabular}
}
\label{parameters}
\end{table}

\noindent The use of MWMF requires an initial set of $M$ measurements, and for each measurement ($m=1,2,\dots,M$) the information regarding cardinality, type and positions of objects obstructing the $m$-th Tx-Rx direct path, indicated as the set of topological parameters $\{\mathcal{T}_{m}\}$. The measurements are used in order to estimate the set of propagation parameters $\{\mathcal{S}\}$ characterizing the model. In this paper, a least square fitting procedure that minimizes the difference between RSS measurements and predictions was adopted to estimate $\{\mathcal{S}\}$. The optimal $\{\mathcal{S}\}_\msf{opt}$ is thus obtained as follows:

\begin{equation}
\label{fitting}
\{\mathcal{S}\}_{\msf{opt}} =  \underset{\{\mathcal{S}\}}{\operatorname{argmin}} \quad \Big\{ \sum_{m=1}^{M} |\msf{RSS}_{m} - \hat{\msf{RSS}}_{m}|^{2}\Big\},
\end{equation}

\noindent where, for the $m$-th available measurement, $\msf{RSS}_{m}$ and $\hat{\msf{RSS}}_{m}$ are the actual vs. the predicted RSS values at Rx, when considering a Tx emitting a known Effective Isotropic Radiated Power (EIRP) $\msf{W^{EIRP}_{TX}}$ at distance $d_{m}$. $\hat{\msf{RSS}}_{m}$ is computed as follows:

\begin{equation}
\hat{\msf{RSS}}_{m} = \msf{W^{EIRP}_{TX}} - \msf{PL}_{\msf{MWMF}}(d_{m}, \{\mathcal{T}_{m}\}, \{\mathcal{S}\}).
\end{equation}

%

The propagation parameters included in $\{\mathcal{S}\}$ may differ from one MWMF model to the other, and can include parameters characterizing both $\msf{PL}_{\msf{OS}}$ and $A_{\msf{MWMF}}$; the set adopted in this work is defined in Section \ref{framework_MWMF}.
\section{ViFi System Model}
\label{sec:Phases} 
\subsection{Offline Phase}
\label{vifi_off}
ViFi uses the MWMF model for the generation of virtual RP fingerprints. Given a set of $L$ WiFi APs in known positions, measurements in a set of $N^{\msf{r}}$ real RPs are first collected, so that a $L \times 1$  RSS fingerprint $\bm{s}_{n_{1}}$ is associated with the $n_{1}$-th RP. The generic $\bm{s}_{n_{1}}$ component, denoted by ${s}_{l,n_{1}}$, contains the RSS received at the $n_{1}$-th RP from the $l$-th AP, obtained by averaging $q>1$ measurements in order to counteract channel variability. The selection of  $q$ is a compromise between accuracy vs. time and effort devoted to measurements. Since only a subset of the $L$ APs may be detected at the generic RP, the $\bm{s}_{n_{1}}$ components of undetected APs are set to a predefined value reflecting lack of detection.\\
The MWMF model is then calibrated on the set of real fingerprints, and used for the generation of $N^{\msf{v}}$ virtual fingerprints associated with virtual RPs. The component $\hat{s}_{l,n_{2}}$ of the generic $L \times 1$ fingerprint $\hat{\bm{s}}_{n_{2}}$ contains the predicted RSS at the $n_{2}$-th virtual RP from the $l$-th AP.
\subsection{Online Phase} 
\label{vifi_on}
ViFi adopts a deterministic W$k$NN estimator using combination of real vs. virtual RPs, to infer target location.\\ 
Denoting by $N=N^{\msf{r}}+N^{\msf{v}}$ the number of RPs in the area $\mathcal{A}$, $\bm{s}_{n}$ ($n=1,2,\dots,N)$ the RSS fingerprint of $n$-th RP, and  $\bm{s}_{i}$ the RSS fingerprint collected during the $i$-th positioning request by a target device in unknown position $\bm{p}_{i}=(x_{i}, y_{i}, z_{i})$, position estimation relies on the computation of a similarity metric $sim_{n,i}=sim(\bm{s}_{n}, \bm{s}_{i})$. The W$k$NN algorithm selects the $k$ RPs that present the highest $sim_{n,i}$ values and provides an estimate of $\bm{p}_{i}$ defined as:

\begin{equation}
\label{wknn_eq}
\hat{\bm{p}}_{i} = \frac{\sum_{n=1}^{k} (sim_{n,i})\bm{p}_{n}}{\sum_{n=1}^{k}sim_{n,i}},
\end{equation}

\noindent where $\bm{p}_{n}=(x_{n}, y_{n}, z_{n})$ is the position of the $n$-th RP in a 3D coordinate system, and $\hat{\bm{p}}_{i}=(\hat{x}_{i}, \hat{y}_{i}, \hat{z}_{i})$ is the estimated position of the target device.\\
$sim_{n,i}$ can be any deterministic metric computable in the RSS space between vectors $\bm{s}_{n}$ and $\bm{s}_{i}$. A popular choice is the inverse Minkowski distance of order $o$, defined as follows:

\begin{equation}
\label{sim}
sim_{n,i} = [\mathcal{D}^{o}_{n,i}]^{-1} = \left[\left(\sum_{l=1}^{L}|s_{l,i}-s_{l,n}|^{o}\right)^{\!\!\frac{1}{o}}\right]^{-1}, \quad o \ge 1.
\end{equation}

Typical orders are  $o=1$ (inverse Manhattan distance) and  $o=2$ (inverse Euclidean distance, used in ViFi). Similarity metrics using modified versions of the inner product between RSS vectors have also been proposed \cite{Torres:2015}, \cite{Caso:Sens:2015}.
\subsection{Offline Phase Implementation Options}
\label{vifi_off_op}
The ViFi offline phase must address two main issues: a) how to use the real RPs, and b) how to determine the number and positions of virtual RPs, as discussed below.  
\subsubsection{Handling real RPs}
\label{subsec:real}
Both cardinality and positions of real RPs are expected to affect the generation of virtual RPs and the resulting ViFi positioning accuracy. Given $N^{\msf{r}}$ real RPs regularly spaced over a grid in $\mathcal{A}$ with area $|\mathcal{A}|$, the spatial RP density is:

\begin{equation}
\label{real_density}
d^{\msf{r}} = \frac{N^{\msf{r}}}{|\mathcal{A}|}.
\end{equation}
 
In \cite{Caso:2017}, \cite{Caso:Fab:2015}, several strategies were proposed for deriving the MWMF propagation parameters; among them, the following two strategies were adopted in this paper:
\begin{description}[style=unboxed, leftmargin = 0cm]
\item [Strategy I: Environment Fitting] -- The set of measurements from all APs to all RPs is used in a global optimization procedure, leading to a common set of propagation parameters for all APs in $\mathcal{A}$. The underlying assumption is that one set of parameters can globally characterize the environment.
\item [Strategy II: Specific AP Fitting] --  The set of measurements from a specific AP to all RPs is used to estimate propagation parameters for that AP. The procedure leads thus to a different set of propagation parameters for each AP.
\end{description}    
\subsubsection{Handling virtual RPs}
\label{subsec:virtual}
In analogy to Equation (\ref{real_density}), denoting with $N^{\msf{v}}$ the number of virtual RPs to be generated, the virtual RPs spatial density is defined as follows:

\begin{equation}
\label{virtual_density}
d^{\msf{v}} = \frac{N^{\msf{v}}}{|\mathcal{A}|}.
\end{equation}

Moreover, positions of virtual RPs in the area can be freely defined: possible options include placement on a grid, as a natural extension of the grid scheme widely adopted for real RPs, and random placement.
\subsection{Online Phase Implementation Options}
\label{vifi_on_op}
The W$k$NN algorithm foresees the selection of two parameters: $k$ and weighting function $sim_{n,i}$. In ViFi the weighting function $sim_{n,i}$ is set to the inverse Euclidean distance while parameter $k$ is selected according to a new procedure. 
Previous work addressed the impact of $k$ on positioning accuracy of deterministic W$k$NN algorithms in real fingerprinting systems, and two approaches emerged: a) a dynamic $k$ selection i.e. that, however, increases complexity without guaranteeing the optimal performance in all cases \cite{Caso:Sens:2015}, \cite{Caso:ICC:2015}, and b) a static $k$ selection, that minimizes the average positioning error over a set of Test Points. The parameter $k$ that minimizes the average error is referred to as $k_{\msf{opt}}$. As shown in \cite{Honkavirta:2009}, \cite{Bahl:2000}, \cite{Kessel:2011}, \cite{Li:2006}, \cite{Torres:2015}, \cite{Caso:ICC:2015}, in real fingerprinting systems $k_{\msf{opt}}$ typically lies in the range between $2$ and $10$.\\
This paper investigates the determination of $k_{\msf{opt}}$ on the basis of system parameters, moving from the assumption that $k_{\msf{opt}}$ depends in particular on the density of real and virtual RPs, and can be thus determined as $k_{\msf{opt}}=\msf{f}(d^{\msf{r}}, d^{\msf{v}})$. In this work a linear law is proposed for $\msf{f}(.,.)$, leading to an approximate $k_{\msf{opt}}$ value, $k_{\msf{est}}$, as follows:

\begin{equation}
\label{k}
k_{\msf{opt}} \approx k_{\msf{est}} = \lceil\alpha(d^{\msf{r}} + d^{\msf{v}})|\mathcal{A}|\rceil, \quad \alpha\ll1.
\end{equation}

Equation (\ref{k}) assumes the same linear dependency of $k_{\msf{opt}}$ on both $d^{\msf{r}}$ and $d^{\msf{v}}$, although virtual fingerprints are different from real ones, since they show a high spatial correlation, with RSS variation between two fingerprints directly related to the physical distance between the corresponding RPs, without the abrupt changes introduced in real fingerprints by measurement errors and channel fading. Two reasons justify this choice: first, there is no model in the literature providing the value of $k_{\msf{opt}}$ as a function of $d^{\msf{r}}$; second, virtual fingerprinting only makes sense if $d^{\msf{r}}$ is much smaller than $d^{\msf{v}}$, thus leading to a low impact of $d^{\msf{r}}$ on $k_{\msf{opt}}$.\\
In general, the values of $\alpha$, $d^{\msf{r}}$ and $d^{\msf{v}}$ to be used in Equation (\ref{k}) will be environment-specific, and will still require a testing phase to determine their exact values. If however a set of values for these parameters valid across different environments can be determined, this will allow to put the system in operation without the need for a testing phase. This possibility is investigated later in this work, first by assessing the validity of Equation (\ref{k}) with the procedure described in Section \ref{framework_kest}, and then by determining and comparing the values of $\alpha$, $d^{\msf{r}}$ and $d^{\msf{v}}$ that lead to the best approximation of $k_{\msf{opt}}$ in the adopted testbeds.
\section{Experimental Analysis Setup}
\label{testbeds}
This section describes the SPinV and TWIST testbeds, implemented at Sapienza University of Rome and Technical University of Berlin, respectively. These two environments were adopted in the initial testing of the ViFi system\footnote{Description of the w-iLab.t I testbed \cite{Lemic:2016}, \cite{VanHaute:2015}, implemented at iMinds in Ghent, Belgium, and used to further assess the ViFi performance, is reported in \cite{ViFi:supplemental}, together with the obtained results.}. In the present section, the procedures defined for the experimental analysis of ViFi are also introduced.
\subsection{SPinV}
\label{SPinV}
Supporting People indoor: a navigation Venture (SPinV) is the indoor positioning testbed implemented at the DIET Department of Sapienza University of Rome. SPinV is deployed in an office environment and covers two floors with an area of approximately $42 \times 12$ $\msf{m}^2$ each. $L_{1}=6$ APs working @ 2.4 and 5 GHz, and $L_{2}=7$ APs working @ 2.4 GHz, with a beacon transmission period of $T_{\msf{b}}=100$ $\msf{ms}$ and a transmit power $\msf{W^{EIRP}_{TX}}=20$ $\msf{dBm}$, are placed at known positions at the $1$-st and $2$-nd floor, respectively. In this work, the SPinV $2$-nd floor has been adopted as evaluation area $\mathcal{A}$, and the APs on this floor have been considered in the fingerprinting measurement campaign, so that $L = L_{2}$.\\ 
Two different measurement campaigns were carried out, each corresponding to a different application scenario:
\begin{description}[style=unboxed, leftmargin = 0cm]
\item [Controlled scenario] -- In this scenario, during the offline phase, $N^{\msf{r},\msf{tot}} = 72$ RPs were selected for fingerprints collection on a regular grid within $\mathcal{A}$; fingerprints were also collected in a set of $N^{\msf{t}}=31$ Test Points (TPs) randomly distributed over $\mathcal{A}$. All measurements were carried out during weekend afternoons, in order to mitigate the effect of radio interference due to possible mobile and temporary connection points. No particular mitigation strategy was instead applied as regards the interference with other APs available in the area. Weekend campaigns also allowed to minimize the variation of propagation conditions due to human presence, considering that no other person than the single human surveyor was present during the RSS collection. A MacBook Pro equipped with an AirPort Extreme Network Interface Card was used, placed on a wooden platform in order to rule out the impact of the human body presence on measured RSS. Both RPs and TPs fingerprints were obtained as the average of $q=50$ scans at each location, in order to remove as much as possible fluctuations due to channel and measurement variability. The RSS collection required approximately $6$ minutes for each RP location.
\item [Crowdsourcing-like scenario] -- In this scenario, $N^{\msf{r},\msf{tot}} = 69$ RPs and  $N^{\msf{t}}=26$ TPs were collected in the area\footnote{Crowdsourced RPs/TPs have the same positions of Controlled scenario RPs/TPs; a few measurements are missing due to practical constraints encountered during the collection campaigns.}. All measurements were performed during weekdays, by a single human surveyor equipped with an Android-based Samsung tablet. In this case, no radio and human interference mitigation strategies were adopted, in order to mimic a realistic crowdsourcing scenario. Normal human office activity (several static and walking persons randomly distributed in the environment) was observed during the RSS collection. RPs fingerprints were obtained by averaging $q=5$ measurements, while TPs fingerprints include a single measurement. 
\end{description} 
A specific characteristic of the SPinV testbed is the location of APs. On both floors, APs are placed in the false ceiling of the central corridor, as shown in Figure \ref{fig:DIET_1_cl}. Furthermore, not all APs cover the entire floor. SPinV can be thus considered an example of sub-optimal environment in terms of signal coverage and APs topology, leading to low spatial diversity.
\subsection{TWIST}
\label{twist}
The TKN Wireless Indoor Sensor network Testbed (TWIST) is the indoor positioning testbed implemented at the TKN headquarter at TUB, in the context of the EVARILOS Project \cite{VanHaute:2015}, \cite{Lemic:2014}. TWIST is deployed in an office environment and covers one floor with an area of about $30 \times 15$ $\msf{m}^2$. Within this environment, $L=4$ dedicated WiFi were placed at known positions and configured to operate @ 2.4 GHz with a beacon transmission period of $T_{\msf{b}}=100$ $\msf{ms}$ and a transmit power $\msf{W^{EIRP}_{TX}}=20$ $\msf{dBm}$. The entire floor was adopted as evaluation environment $\mathcal{A}$, and all the $L$ APs were included in the fingerprinting measurement campaign. During the fingerprinting offline phase, $N^{\msf{r},\msf{tot}} = 41$ RPs were identified on a regular grid and measured within $\mathcal{A}$; an additional set of $N^{\msf{t}}=80$ uniformly distributed TPs was collected.\\
In the case of TWIST, all measurements were taken with the same settings defined for the Controlled scenario in SPinV, and no Crowdsourcing-like scenario was considered\footnote{RSS data collected in SPinV and TWIST testbeds, as well as in w-iLab.t I analyzed in \cite{ViFi:supplemental}, are available for download at: please contact the authors.}.\\ 
Oppositely to SPinV, APs in TWIST are approximately placed at the four corners of the area, as shown in Figure \ref{fig:TWIST_1_cl}. TWIST can be thus considered as an optimal environment in terms of APs topology, characterized by high signal coverage and spatial diversity. 

\subsection{Procedure for the Analysis of the MWMF Accuracy in Virtual RPs Generation}
\label{framework_MWMF}
In agreement with the approach proposed in \cite{Caso:2017}, \cite{Caso:Fab:2015}, the following procedure  was adopted for the analysis of the reliability of the MWMF model in generating virtual RPs:
\begin{enumerate}
\item A parameter $\rho$ was defined, in order to determine the number of RPs, $N^{\msf{r}}$, used for the model fitting reported in Equation (\ref{fitting}) of Section \ref{sec:MWMF}, out of the total number of RPs, $N^{\msf{r},\msf{tot}}$, so that $N^{\msf{r}} = \lceil \rho N^{\msf{r},\msf{tot}} \rceil$. The accuracy in estimating the propagation parameters and generating virtual RPs was then evaluated as a function of $\rho$. Figure \ref{fig:DIET_1_cl}  shows the positions of the $N^{\msf{r},\msf{tot}}$  RPs, of the APs  and of the  $N^{\msf{r}}$ RPs selected for model fitting as a function of $\rho$ in SPinV, while Figure \ref{fig:TWIST_1_cl} provides the same information for TWIST.
\begin{figure}[!h]
\centering
\includegraphics[trim=0 1cm 0 0cm, width=3.7in]{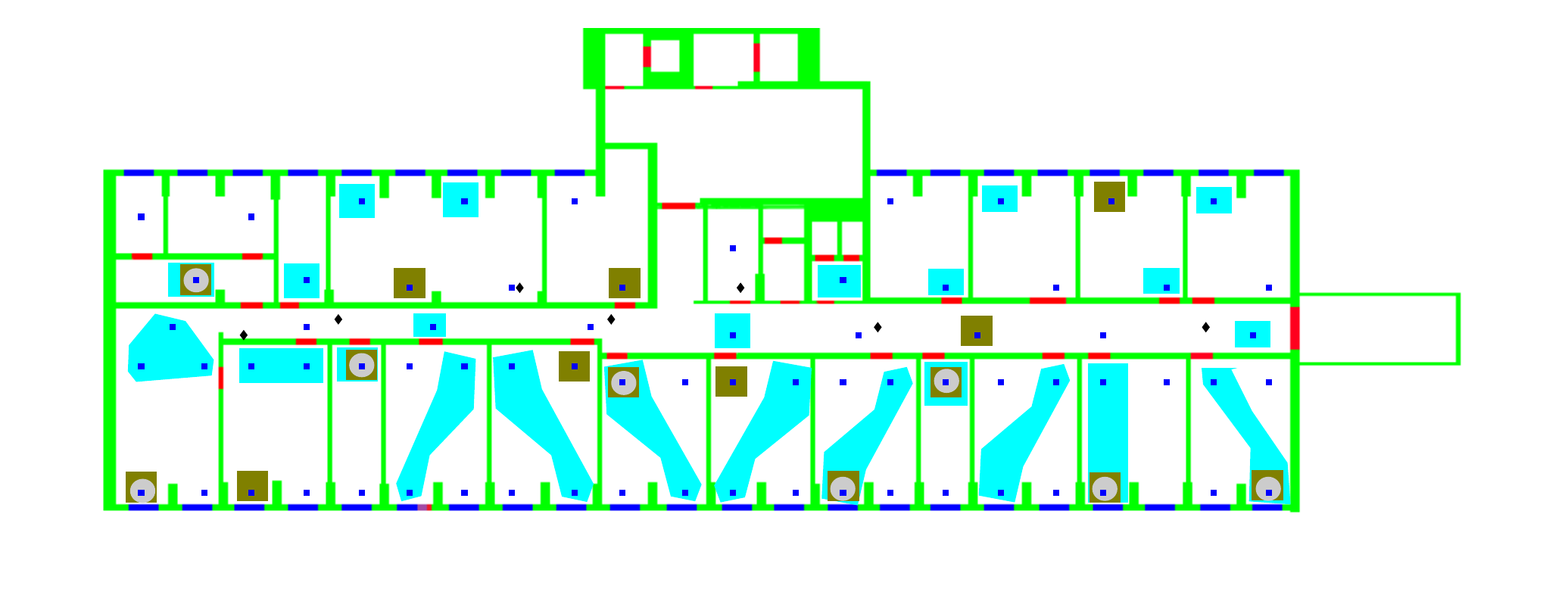}
\caption{Position of real RPs  (blue squares) and APs (black diamonds) in the SPinV testbed. Areas of different colors highlight selected RPs when $\rho=0.1$ (light grey), $\rho=0.2$ (dark green), $\rho=0.5$ (light blue), $\rho=1$ (white $+$ all colors).}
\label{fig:DIET_1_cl}
\end{figure}
\begin{figure}[!h]
\centering
\includegraphics[trim=0 1cm 0 0cm, width=3in]{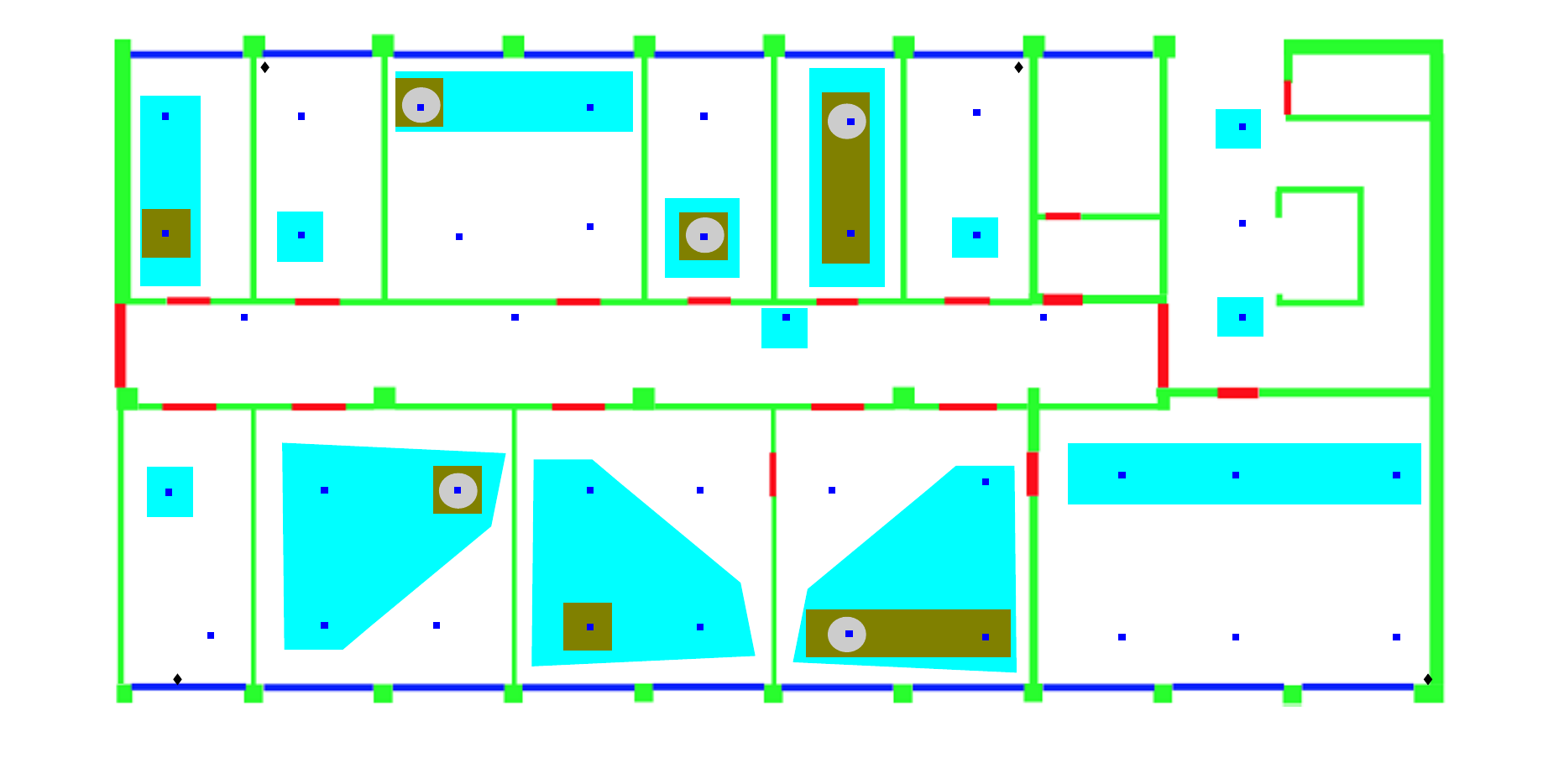}
\caption{Position of real RPs  (blue squares) and APs (black diamonds) in the TWIST testbed. Areas of different colors highlight selected RPs when $\rho=0.1$ (light grey), $\rho=0.2$ (dark green), $\rho=0.5$ (light blue), $\rho=1$ (white $+$ all colors).}
\label{fig:TWIST_1_cl}
\end{figure}
\item The model fitting procedure defined by Equation (\ref{fitting}), was carried out to evaluate propagation parameters to be used in the MWMF model for each of the two selection strategies introduced in Section \ref{subsec:real}. TABLE \ref{model_params} defines the set $\{\mathcal{S}\}$ of propagation parameters included in the MWMF model fitting procedure, and the settings for the remaining parameters in Equation (\ref{fitting}).
\begin{table}[!ht]
\centering
\caption{Model parameters setting.}
\begin{tabular}{p{0.1\textwidth}c}
\toprule
\textbf{Parameter} & \textbf{Setting} \\
\midrule
$N_{\msf{obj}}$ & $2$ (walls, doors)\\
\midrule
$I_{\msf{walls}}$ & $1$\\
\midrule
$I_{\msf{doors}}$ & $1$\\
\midrule
$N_{\msf{f}}$ & $0$ \\
\midrule
$\{\mathcal{S}\}$ & $\{\gamma, l_{c}, l_{\msf{wall}}, l_{\msf{door}}\}$ \\
\bottomrule
\end{tabular}
\label{model_params}
\end{table}
\item For each selection strategy, RSS values were predicted, using the MWMF model, in the same locations where the $N^{\msf{r},\msf{tot}}$ RPs were originally collected. Next, the accuracy of the model was evaluated by defining the prediction error as follows:
\begin{equation}
\delta_{l,n} (\rho) = |s_{l,n} - \hat{s}_{l,n} (\rho)|,
\end{equation}
where $\hat{s}_{l,n} (\rho)$ is the predicted RSS for the generic ($\msf{AP}_{l}$, $\msf{RP}_{n}$) pair, obtained by using a set of $N^{\msf{r}} = \lceil \rho N^{\msf{r},\msf{tot}} \rceil$ RPs in the model fitting procedure, while $s_{l,n}$ is the measured RSS for the same pair. Assuming the generic $\delta_{l,n} (\rho)$ value as a sample of a random variable $\delta_l (\rho)$ related to the $l$-th AP, the cumulative distribution function (CDF) of $\delta_l (\rho)$ and the average error $\bar{\delta_l} (\rho) = \frac{\sum_{n=1}^{N^{\msf{r},\msf{tot}}}\delta_{l,n} (\rho)}{N^{\msf{r},\msf{tot}}}$ were also evaluated.
\end{enumerate}
Results of the analysis are reported in Section \ref{results_estimation} and \ref{results_crowdsourcing} for Controlled and Crowdsourcing-like scenarios, respectively.
\subsection{Procedure for the Analysis of the Impact of Virtual RPs on Positioning Accuracy}
\label{framework_accuracy}
The performance of ViFi was evaluated as a function of: a) the density $d^{\msf{r}}$ and the selection strategy of real RPs, b) the density $d^{\msf{v}}$ and  placement of virtual RPs, and c) the $k$ parameter in the W$k$NN algorithm. The analysis was carried out by computing the positioning error $\epsilon_i (d^{\msf{r}}, d^{\msf{v}}, k)$ for each TP $i$ ($i=1,2,\dots,N^{\msf{t}}$) as follows:   
\begin{equation}
\label{epsi}
\epsilon_{i}(d^{\msf{r}}, d^{\msf{v}}, k) = \sqrt{(x_{i}-\hat{x}_{i})^2 + (y_{i}-\hat{y}_{i})^2 + (z_{i}-\hat{z}_{i})^2},
\end{equation}
\noindent where $(x_{i}, y_{i}, z_{i})=\bm{p}_{i}$ and $(\hat{x}_{i}, \hat{y}_{i}, \hat{z}_{i})=\hat{\bm{p}}_{i}$ are the actual and the estimated positions of the $i$-th target device, respectively. Note that dependence of $\hat{\bm{p}}_{i}$ components on the $\left\lbrace d^{\msf{r}}, d^{\msf{v}}, k\right\rbrace$ set was omitted in Equation (\ref{epsi}) for the sake of readability. As in the case of the prediction error $\delta_{l} (\rho)$, the CDF of positioning error $\epsilon (d^{\msf{r}}, d^{\msf{v}}, k)$ and the average positioning error $\bar{\epsilon} (d^{\msf{r}}, d^{\msf{v}}, k) = \frac{\sum_{i=1}^{N^{\msf{t}}}\epsilon_{i}(d^{\msf{r}}, d^{\msf{v}}, k)}{N^{\msf{t}}}$ were evaluated. Furthermore, a performance indicator, referred to as \emph{virtualization} gain $\mathcal{G}(d^{\msf{r}}, d^{\msf{v}}, k)$, was introduced to measure the gain in accuracy achieved when using a density of virtual RPs $d^{\msf{v}}>0$. The virtualization gain $\mathcal{G}(d^{\msf{r}}, d^{\msf{v}}, k)$ is defined as follows:
\begin{equation}
\label{gain_def}
\mathcal{G}(d^{\msf{r}}, d^{\msf{v}}, k) = \frac{\bar{\epsilon} (d^{\msf{r}}, 0, k)}{\bar{\epsilon} (d^{\msf{r}}, d^{\msf{v}}, k)}, \quad d^{\msf{v}}>0.
\end{equation}
Results of the analysis are reported in Section \ref{results_accuracy} and \ref{results_crowdsourcing} for Controlled and Crowdsourcing-like scenarios, respectively.
\subsection{Procedure for testing the validity of the  $k_{\msf{est}}$ model}
\label{framework_kest} 
The following procedure was applied in order to assess the validity of Equation (\ref{k}) in both SPinV and TWIST testbeds:
\begin{enumerate}
\item Given a generic $(d^{\msf{r}}, d^{\msf{v}})$ combination,  $k_{\msf{opt}}$ and in turn $\bar{\epsilon}(d^{\msf{r}}, d^{\msf{v}}, k_{\msf{opt}})$ were evaluated;
\item For each $\alpha \in [\alpha_{\msf{min}}:\alpha_{\msf{max}}]$, $k_{\msf{est}}(\alpha)$ was evaluated following Equation (\ref{k}), and the corresponding average positioning error $\bar{\epsilon}(d^{\msf{r}}, d^{\msf{v}}, k_{\msf{est}}(\alpha))$ was determined;
\item The difference between $\bar{\epsilon}(d^{\msf{r}}, d^{\msf{v}}, k_{\msf{est}}(\alpha))$ and $\bar{\epsilon}(d^{\msf{r}}, d^{\msf{v}}, k_{\msf{opt}})$ was evaluated and denoted with $\beta (d^{\msf{r}}, d^{\msf{v}}, \alpha)$:
\begin{equation}
\label{beta}
\beta (d^{\msf{r}}, d^{\msf{v}}, \alpha) = \bar{\epsilon}(d^{\msf{r}}, d^{\msf{v}}, k_{\msf{est}}(\alpha)) - \bar{\epsilon}(d^{\msf{r}}, d^{\msf{v}}, k_{\msf{opt}})
\end{equation}
\end{enumerate} 
Numerical results of the above procedure for both SPinV and TWIST testbeds are reported in Section \ref{results_kest}.\\
TABLE \ref{exp_settings} presents the values for the parameters introduced in the procedures defined in the current section, as well as in Sections \ref{framework_MWMF} and \ref{framework_accuracy}.

\begin{table}[!ht]
\centering
\caption{Experimental parameters setting.}
\resizebox{\columnwidth}{!}{
\begin{tabular}{p{0.1\textwidth}cc}
\toprule
\textbf{Parameter} & \textbf{SPinV} & \textbf{TWIST} \\
\midrule
$\{\rho\}$ & \multicolumn{2}{c}{$\{0.1, 0.2, 0.5, 1\}$}\\
\midrule
$\{d^{\msf{r}}\}$ & $\{0.02, 0.03, 0.07, 0.14\}$  & $\{0.01, 0.02, 0.05, 0.09\}$ \\
\midrule
$\{N^{\msf{r}}\}$ & $\{8, 15, 36, 72\}$ & $\{5, 9, 21, 41\}$ \\
\midrule
$\{d^{\msf{v}}\}$ & \multicolumn{2}{c}{$\{0.01, 0.05, 0.1, 0.5, 1, 5, 10 \}$}\\
\midrule
$\{N^{\msf{v}}\}$ & $\{6, 26, 51, 252, 504, 2520, 5040\}$ & $\{5, 23, 45, 225, 450, 2250, 4500\}$ \\
\midrule
$\{\alpha_{\msf{min}}, \alpha_{\msf{max}}\}$ & \multicolumn{2}{c}{$\{0.01, 0.25\}$}\\
\bottomrule
\end{tabular}
}
\label{exp_settings}
\end{table}

\section{Experimental Results and Discussion}
\label{results}
\subsection{Controlled Scenario: Accuracy in Generation of Virtual RPs}
\label{results_estimation}
\begin{figure}[!h]
\centering
\subfloat[][\label{delta1_S1_SPinV}]
{\includegraphics[trim=0 0 0 0, width=1.68in]{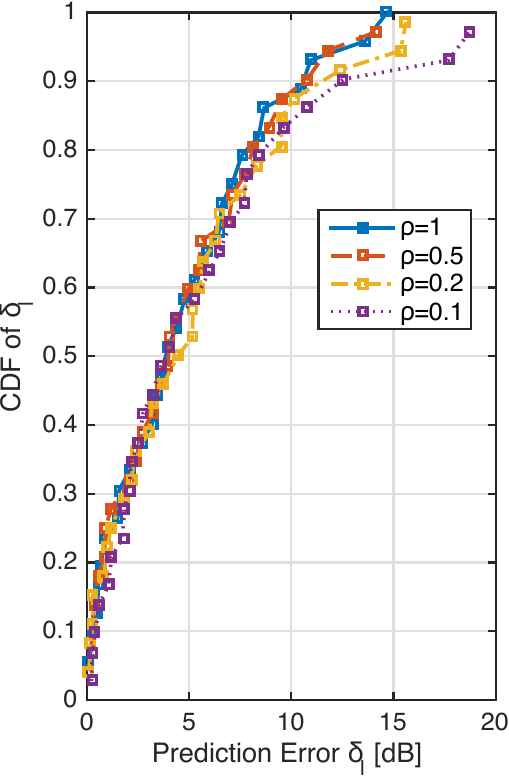}} \quad
\subfloat[][\label{delta1_S2_SPinV}]
{\includegraphics[trim=0 0 0 0, width=1.68in]{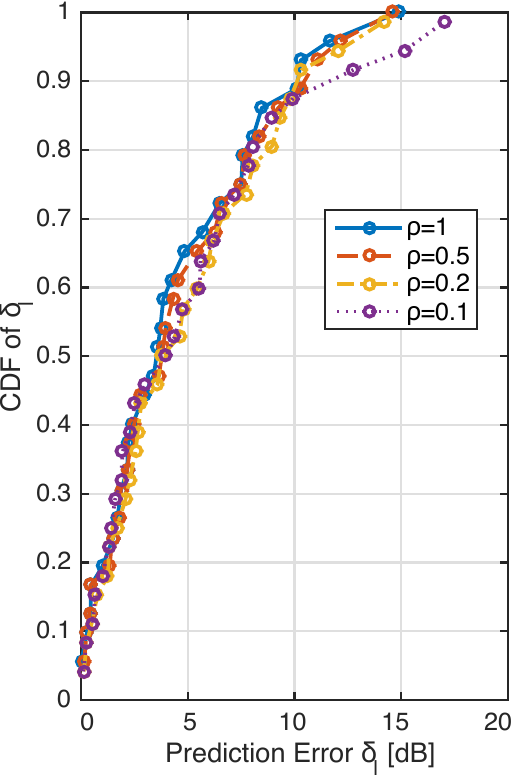}} \\
\caption{Cumulative Distribution Function of prediction error $\delta_l(\rho)$ for a reference AP in SPinV. Strategy I (a) vs. Strategy II (b).}
\label{delta_1_SPinV}
\end{figure}

\begin{figure}[!h]
\centering
\subfloat[][\label{delta1_S1_TWIST}]
{\includegraphics[trim=0 0 0 0, width=1.68in]{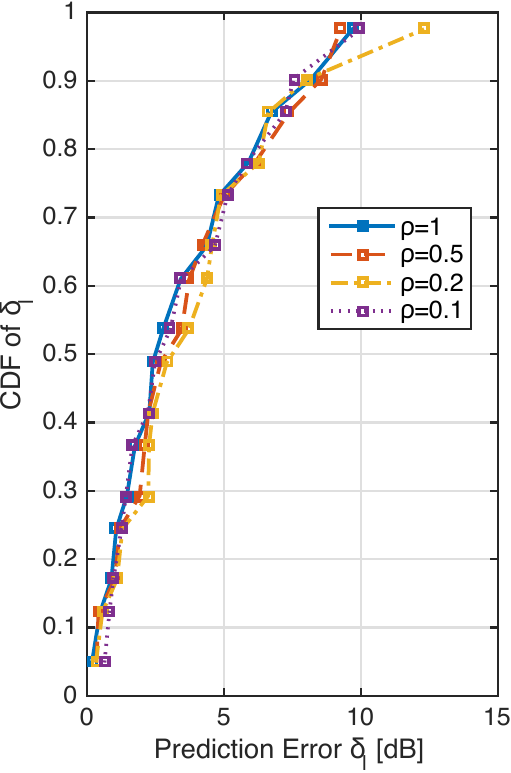}} \quad
\subfloat[][\label{delta1_S2_TWIST}]
{\includegraphics[trim=0 0 0 0, width=1.68in]{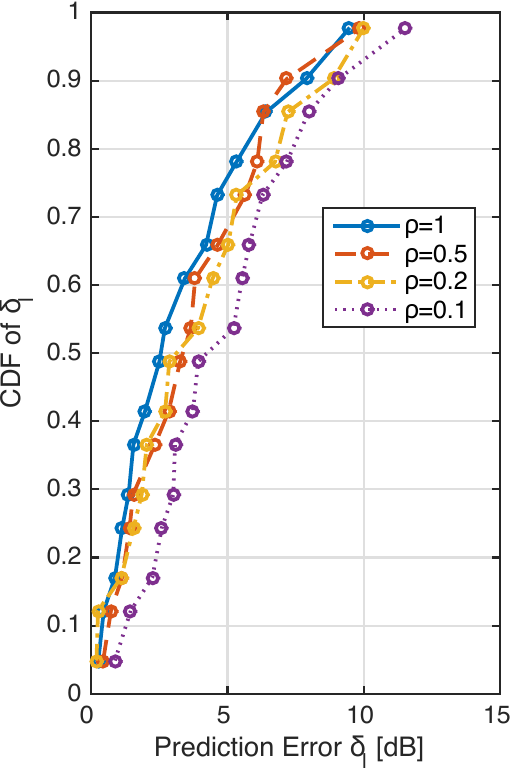}} \\
\caption{Cumulative Distribution Function of prediction error $\delta_l(\rho)$ for a reference AP in TWIST. Strategy I (a) vs. Strategy II (b).}
\label{delta_1_TWIST}
\end{figure}

The MWMF model adopted in ViFi for the generation of virtual RPs was analyzed and compared against the OS model proposed in \cite{Widyawan:2007}, \cite{Pulkkinen:2015}.\\
Figures \ref{delta_1_SPinV} and \ref{delta_1_TWIST} show the CDFs of the prediction error $\delta_l(\rho)$ for selection Strategy I (Figures \ref{delta1_S1_SPinV} and \ref{delta1_S1_TWIST}) and Strategy II (Figures \ref{delta1_S2_SPinV} and \ref{delta1_S2_TWIST}) for a reference AP within the SPinV and TWIST testbeds. Results show that, for both strategies and testbeds, slightly different $\delta_l(\rho)$ errors are obtained as $\rho$ increases from $0.1$ to $1$. Strategies perform similarly for the reference AP in the SPinV environment, with a slight performance improvement given by Strategy II, while, within the TWIST environment, Strategy I allows a faster convergence to low $\delta_l(\rho)$ values with respect to Strategy II.

\begin{figure}
\centering
\subfloat[][\label{Ia_SPinV}]
{\includegraphics[trim=0 0 0 2cm, width=1.68in]{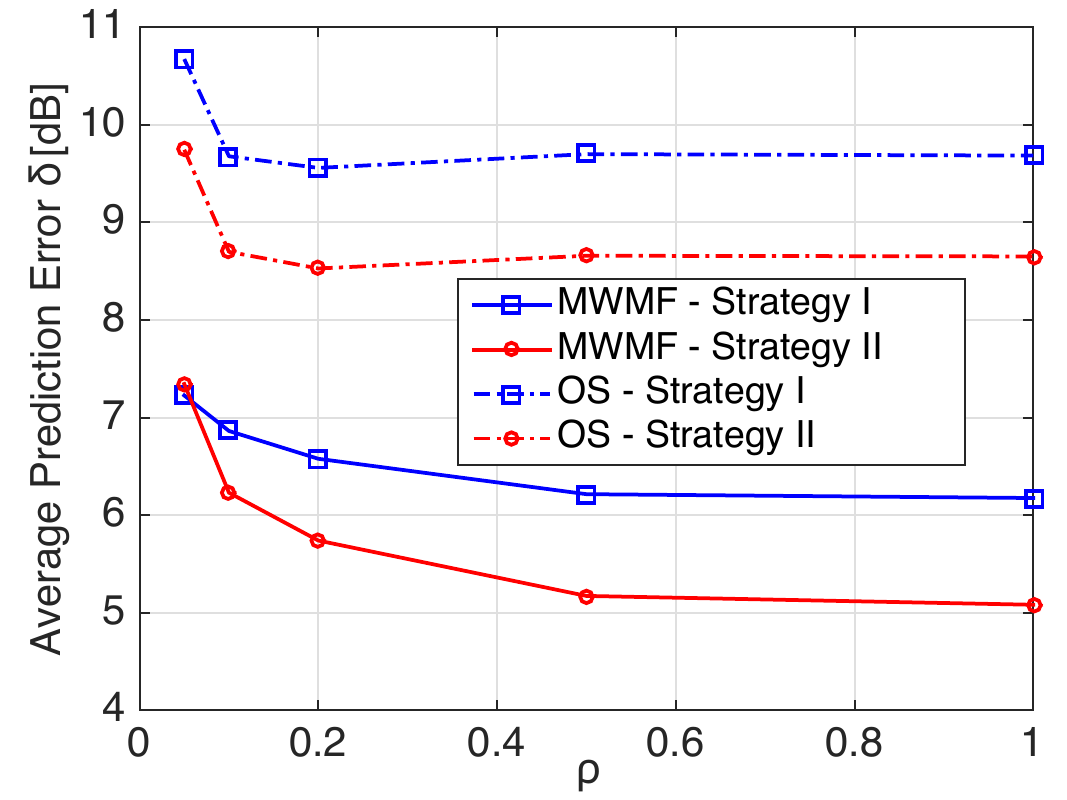}} \quad
\subfloat[][\label{Ia_TWIST}]
{\includegraphics[trim=0 0 0 2cm, width=1.68in]{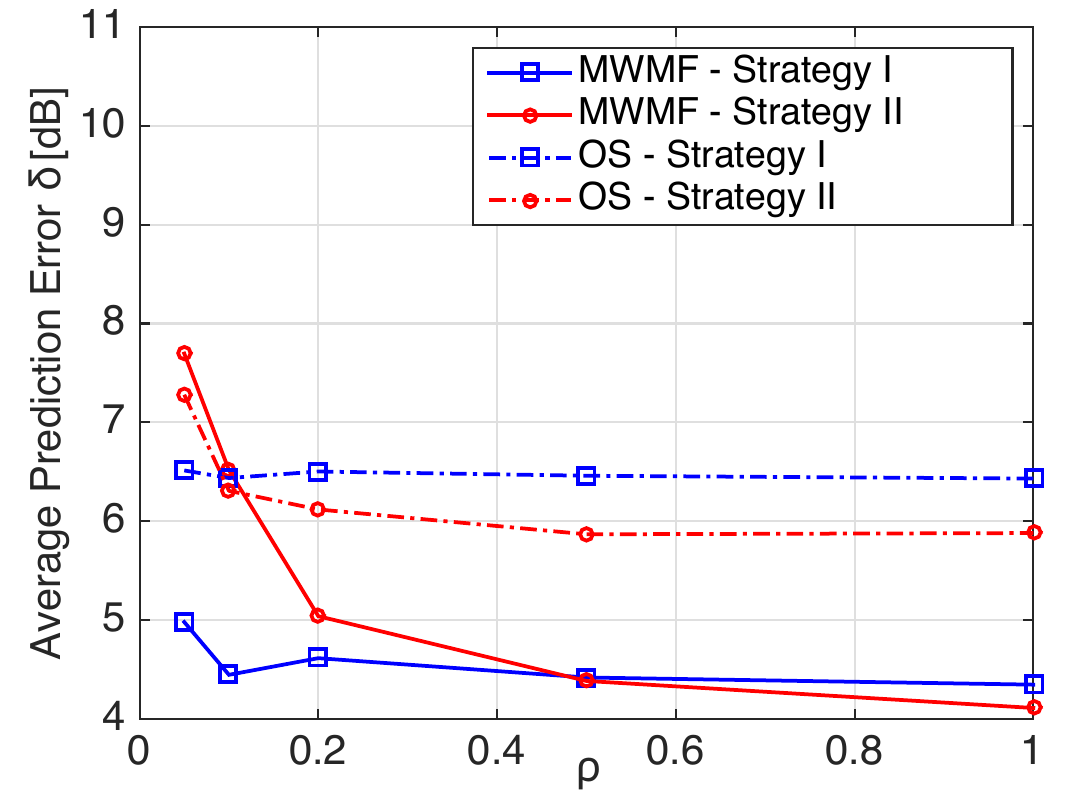}}
\caption{Average prediction error $\bar{\delta}(\rho)$ as a function of $\rho$ for Strategies I and II in SPinV (a) vs. TWIST (b) (Continuous lines: proposed MWMF model; dashed-dotted lines: OS model proposed in \cite{Widyawan:2007}, \cite{Pulkkinen:2015}.)}
\label{delta_m_average_error}
\end{figure}

\noindent Results suggest that, when a relatively large value of $\rho$ (and in turn of $N^{\msf{r}}$ and $d^{\msf{r}}$) is used, Strategy II outperforms Strategy I. Oppositely, when a relatively low $\rho$ value is used, Strategy I achieves slightly better performance than Strategy II, since it combines measurements from different APs, taking advantage of environment homogeneity. This is confirmed by Figure \ref{delta_m_average_error}, showing the average prediction error $\bar{\delta}(\rho)$, obtained by averaging all the $\bar{\delta_l}(\rho)$ values over the $L$ APs, as a function of $\rho$ (in addition to the $\rho$ values in TABLE  \ref{exp_settings}, values of $\bar{\delta}(\rho)$ obtained for $\rho = 0.05$ are also shown), in SPinV and TWIST, respectively. The Figure also presents results obtained using the OS model. Results show that Strategy II is the optimal selection strategy in SPinV, since the initial amount of real RPs is large enough to guarantee a better RSS prediction when specific propagation parameters are associated to different APs. Oppositely, Strategy I provides a better RSS prediction in the TWIST testbed for low values of $\rho$, taking advantage in particular of the symmetry in the AP placement in TWIST. Figure \ref{delta_m_average_error} also shows that the proposed MWMF model outperforms the OS model, with a decrease in the average prediction error of up to $4$ $\msf{dBs}$ in SPinV and $2$ $\msf{dBs}$ in TWIST.
\begin{figure}
\centering
\subfloat[][\label{IIa_SPinV}]
{\includegraphics[trim=0 0 0 0,width=1.68in]{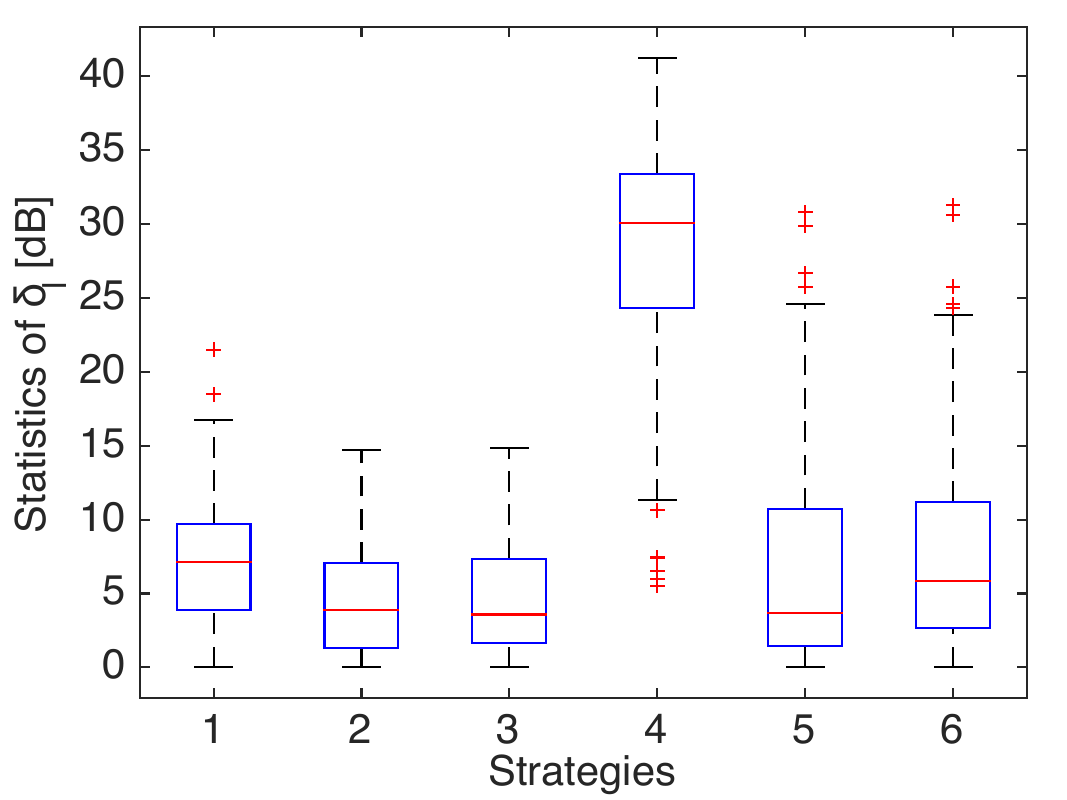}} \quad
\subfloat[][\label{IIa_TWIST}]
{\includegraphics[trim=0 0 0 0,width=1.68in]{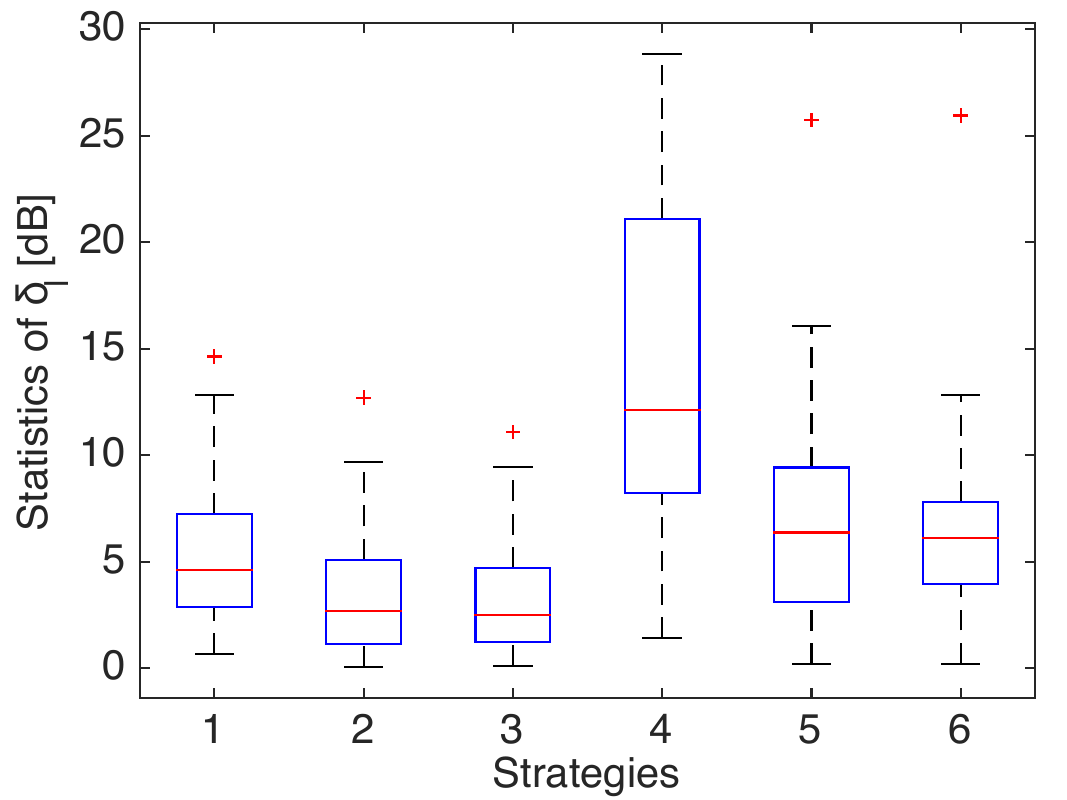}} \\
\caption{Statistics of $\delta_l(\rho)$ for a reference AP and $\rho=1$ in SPinV (a) vs. TWIST (b) (1: MWMF-No Fit; 2: MWMF-Strategy I; 3: MWMF-Strategy II; 4: OS-No Fit; 5: OS-Strategy I; 6: OS-Strategy II). }
\label{delta_boxplot}
\end{figure}
Results presented so far highlight that few, uniformly distributed, measurements are in most cases sufficient to obtain a reliable estimation of the propagation parameters and generation of virtual RPs, supporting the approach proposed in ViFi.\\
One might ask: are measurements required at all? The question is answered by results shown in Figure \ref{delta_boxplot}, presenting the statistics of $\delta_{l}(\rho)$ for a reference AP and $\rho=1$ for MWMF vs. OS models. Each model was fitted according to Strategy I, Strategy II and to a baseline \emph{No Fit} strategy in which no fitting is carried out. For the \emph{No Fit} strategy, the RSS predictions were obtained in MWMF by using propagation parameters estimated for a totally different area $\mathcal{A}^{'}$, reported in \cite{Borrelli:2004}; in the OS model the settings $\gamma=2$, $l_0=20$ were adopted. Note that in Figure \ref{delta_boxplot}, and following ones, error statistics are summarized in the form of boxplot diagrams showing minimum, maximum and median values, $25$-th and $75$-th percentiles, and possible outliers. Results show that the prediction error significantly increases for both MWMF and OS models, and in both testbeds, when no site-specific model fitting is carried out: measurements can be thus significantly reduced but not totally avoided, because of the empirical nature of the models. 



\subsection{Controlled Scenario: Positioning Accuracy}
\label{results_accuracy}
In this section, experimental results on the positioning accuracy of ViFi are presented for both testbeds in the Controlled scenario. Next, results of the application of the empirical procedure presented in Section \ref{framework_kest} for estimating the value of $k_{\msf{opt}}$ in the ViFi online phase are reported.
\subsubsection{Impact of $d^{\msf{r}}$}
\label{results_dr}
\begin{figure}[!h]
\centering
\subfloat[][\label{dr_subplot_SPinV_a}]
{\includegraphics[trim=0 0 0 0,width=1.2in]{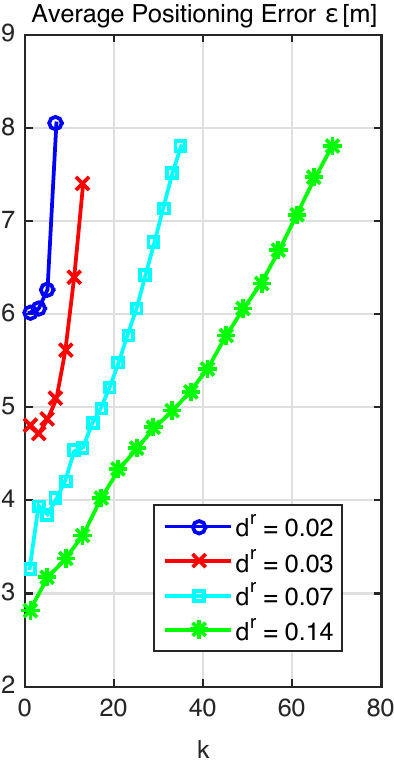}} \quad
\subfloat[][\label{dr_subplot_SPinV_b}]
{\includegraphics[trim=0 0 0 0,width=1.8in]{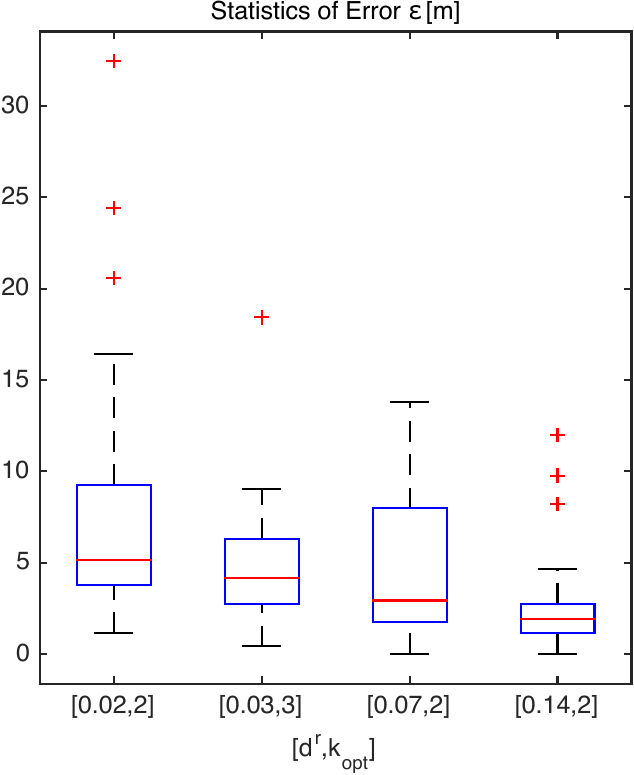}} \\
 \caption{Average positioning error $\bar{\epsilon}(d^{\msf{r}}, 0, k)$ as a function of $k$, for $d^{\msf{r}}$ as in TABLE \ref{exp_settings} (a) and statistics of $\epsilon(d^{\msf{r}}, 0, k_{\msf{opt}})$ (b) in SPinV.}
\label{dr_subplot_SPinV}
\end{figure}

\begin{figure}[!h]
\centering
\subfloat[][\label{dr_subplot_TWIST_a}]
{\includegraphics[trim=0 0 0 0,width=1.2in]{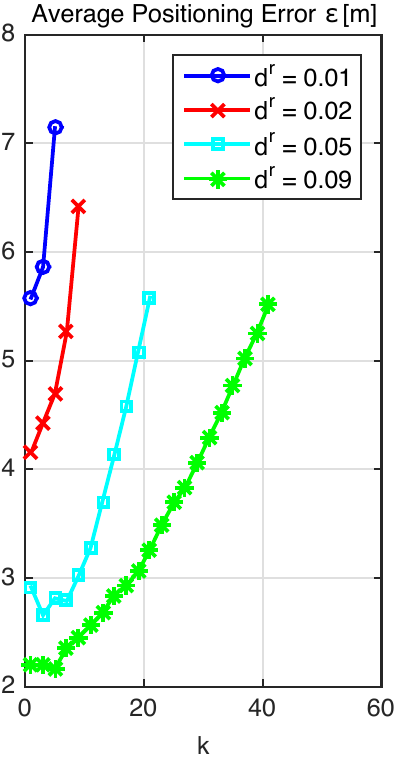}} \quad
\subfloat[][\label{dr_subplot_TWIST_b}]
{\includegraphics[trim=0 0 0 0,width=1.8in]{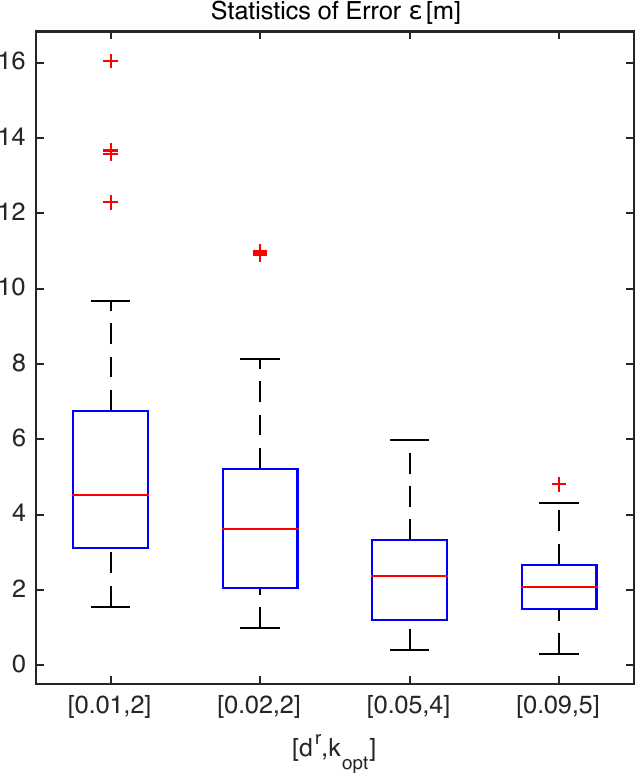}} \\
 \caption{Average positioning error $\bar{\epsilon}(d^{\msf{r}}, 0, k)$  as a function of $k$, for $d^{\msf{r}}$ as in TABLE \ref{exp_settings} (a), and statistics of $\epsilon(d^{\msf{r}}, 0, k_{\msf{opt}})$ (b) in TWIST.}
\label{dr_subplot_TWIST}
\end{figure}

Figures \ref{dr_subplot_SPinV} and \ref{dr_subplot_TWIST} show the impact of the density of real RPs on the average positioning accuracy, when no virtual RPs are introduced, for SPinV vs. TWIST testbeds.\\
Figures \ref{dr_subplot_SPinV_a} and \ref{dr_subplot_TWIST_a} report the average positioning error $\bar{\epsilon}(d^{\msf{r}}, 0, k)$ as a function of $k$ in SPinV vs. TWIST. Results show that the positioning error decreases as $d^{\msf{r}}$ increases. The decrease of $\bar{\epsilon}(d^{\msf{r}}, 0, k)$ is less and less significant as $d^{\msf{r}}$ increases, suggesting the presence of a lower bound, in agreement with \cite{Bahl:2000} and others, possibly due to the inherent measurement error in the collection of real RPs. Results also show that a different lower bound for the error is reached in the two testbeds, about $3$ $\msf{m}$ for SPinV vs. $2$ $\msf{m}$ for TWIST, and indicate that optimal AP placement positively affects the system performance, in agreement with \cite{Baala:2009}.\\
Figures \ref{dr_subplot_SPinV_b} and \ref{dr_subplot_TWIST_b} present the boxplot of the positioning error $\epsilon(d^{\msf{r}}, 0, k_{\msf{opt}})$ as a function of $d^{\msf{r}}$, and confirm that in real fingerprinting systems the value of $k_{\msf{opt}}$ spans in the $[2,10]$ range, as observed in \cite{Honkavirta:2009}, \cite{Bahl:2000}, \cite{Kessel:2011}, \cite{Li:2006}, \cite{Torres:2015}, \cite{Caso:ICC:2015}. 

\subsubsection{Impact of real RPs selection strategies and virtual RPs placement}
\label{results_strategies}
\begin{figure}
\centering
\subfloat[][\label{Ib}]
{\includegraphics[trim=0 0 0 0, width=1.68in]{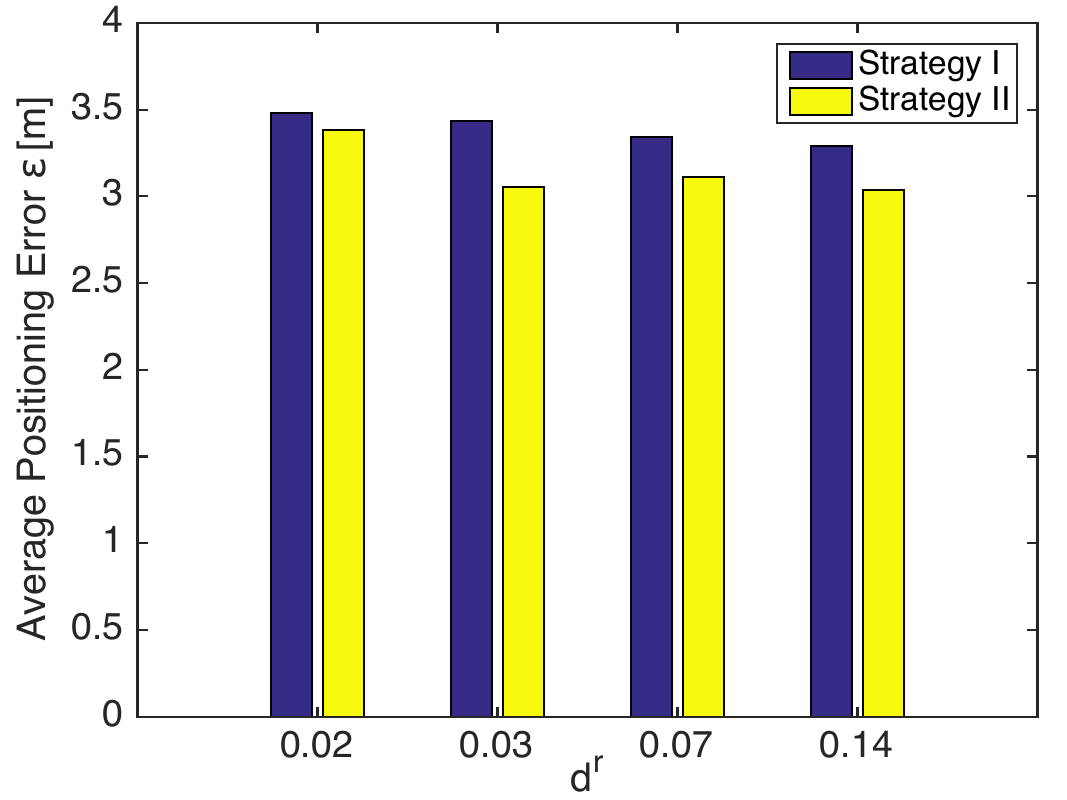}} \quad
\subfloat[][\label{IIb}]
{\includegraphics[trim=0 0 0 0, width=1.68in]{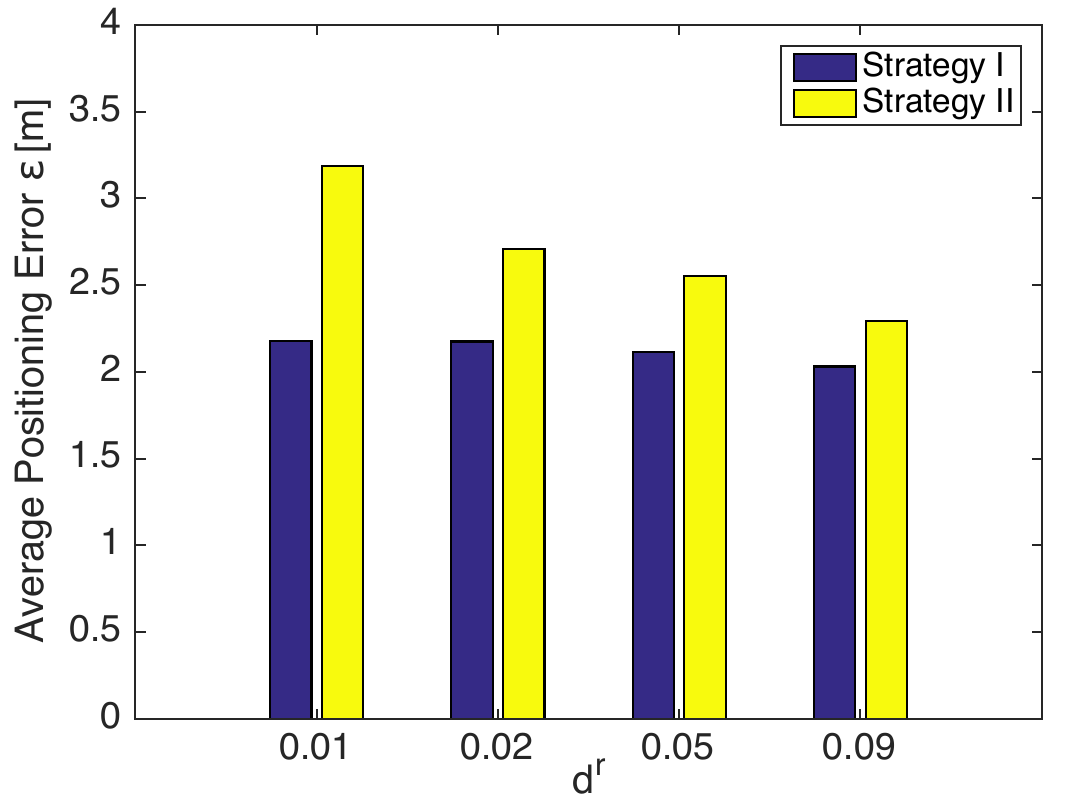}} \\
\caption{Average positioning error $\bar{\epsilon}(d^{\msf{r}}, d^{\msf{v}}_{\msf{max}}, k_{\msf{opt}})$, for $d^{\msf{r}}$ as in TABLE \ref{exp_settings}, and for selection Strategy I and II, in SPinV (a) vs. TWIST (b) (Blue bars: Strategy I; yellow bars: Strategy II).}
\label{IvsII}
\end{figure}

Figure \ref{IvsII} reports the average positioning error $\bar{\epsilon}(d^{\msf{r}}, d^{\msf{v}}_{\msf{max}}, k_{\msf{opt}})$ for Strategy I and II, and $d^{\msf{r}}$ as in TABLE \ref{exp_settings}. Results are in agreement with Figure \ref{delta_m_average_error}: within the SPinV environment, Strategy II outperforms Strategy I, and strategies performance approach when $d^{\msf{r}} = d^{\msf{r}}_{\msf{min}}$; on the contrary, within the TWIST environment, Strategy I leads to better positioning accuracy, and strategies performance approach when $d^{\msf{r}} = d^{\msf{r}}_{\msf{max}}$. In conclusion one can observe that: 1) the selection of the optimal strategy depends on the number of real RPs, and 2) a direct relationship exists between the average prediction error $\bar{\delta}(\rho)$ and the average positioning error $\bar{\epsilon}(d^{\msf{r}}, d^{\msf{v}}, k)$: the strategy minimizing the average prediction error $\bar{\delta}(\rho)$ also minimizes the average positioning error, and should thus be selected.\\
As for the placement of virtual RPs, both grid and random placements were investigated, according to the discussion in Section \ref{subsec:virtual}, with results showing no significant difference between the two placements. As a consequence, a grid placement was adopted in all experimental results shown throughout the paper.

\subsubsection{Impact of $d^{\msf{v}}$}
\label{results_dv}
\begin{figure}[!h]
\centering
\centering
\subfloat[][\label{dv_subplot_1_SPinV_a}]
{\includegraphics[trim=0 0 0 0,width=1.2in]{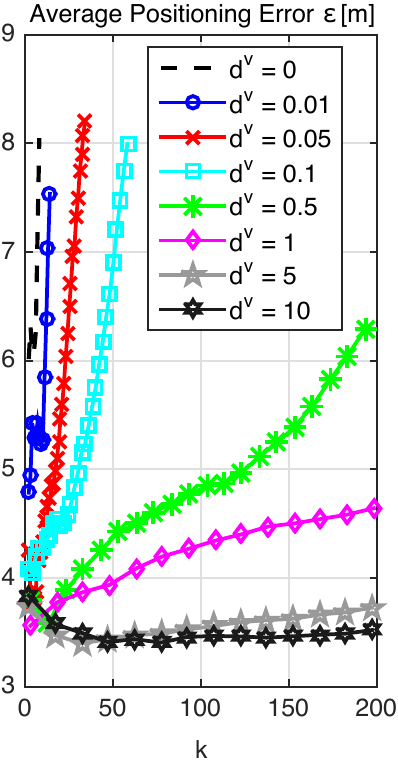}} \quad
\subfloat[][\label{dv_subplot_1_SPinV_b}]
{\includegraphics[trim=0 0 0 0,width=1.9in]{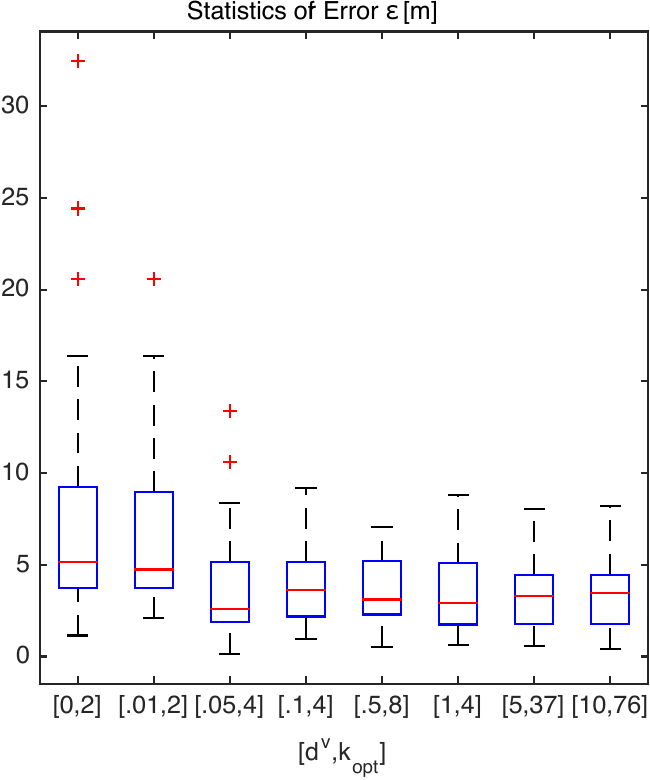}} \\
 \caption{Average positioning error $\bar{\epsilon}(d^{\msf{r}}_{\msf{min}}, d^{\msf{v}}, k)$ as a function of $k$, for $d^{\msf{v}}$ as in TABLE \ref{exp_settings} (a), and statistics of $\epsilon(d^{\msf{r}}_{\msf{min}}, d^{\msf{v}}, k_{\msf{opt}})$ (b) in SPinV.}
\label{dv_subplot_1_SPinV}
\end{figure}

\begin{figure}[!h]
\centering
\subfloat[][\label{dv_subplot_1_TWIST_a}]
{\includegraphics[trim=0 0 0 0,width=1.2in]{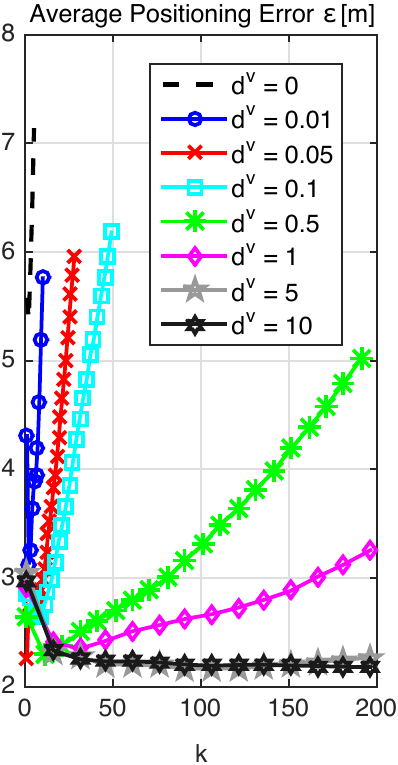}} \quad
\subfloat[][\label{dv_subplot_1_TWIST_b}]
{\includegraphics[trim=0 0 0 0,width=1.9in]{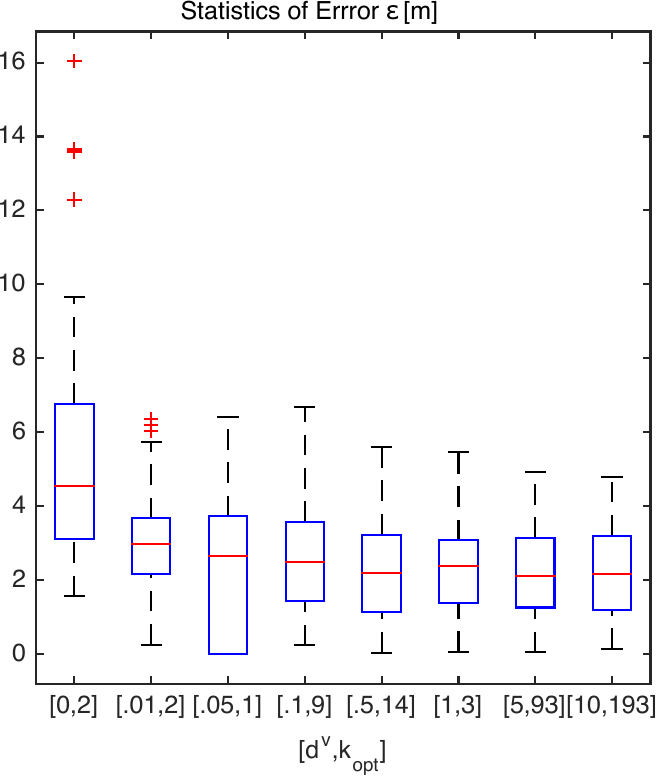}} \\
 \caption{Average positioning error $\bar{\epsilon}(d^{\msf{r}}_{\msf{min}}, d^{\msf{v}}, k)$ as a function of $k$, for $d^{\msf{v}}$ as in TABLE \ref{exp_settings} (a), and statistics of $\epsilon(d^{\msf{r}}_{\msf{min}}, d^{\msf{v}}, k_{\msf{opt}})$ (b) in TWIST.}
\label{dv_subplot_1_TWIST}
\end{figure}

\begin{figure}
\centering
\subfloat[][\label{Ig}]
{\includegraphics[trim=0 0 0 0, width=1.68in]{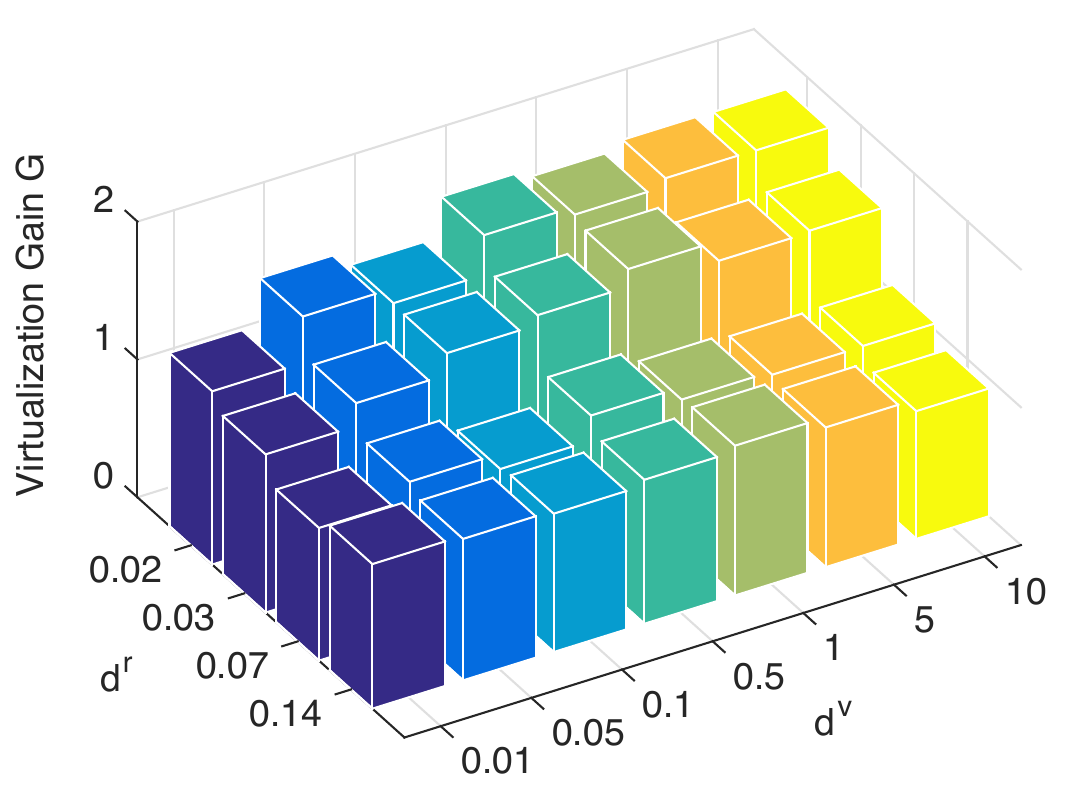}} \quad
\subfloat[][\label{IIg}]
{\includegraphics[trim=0 0 0 0, width=1.68in]{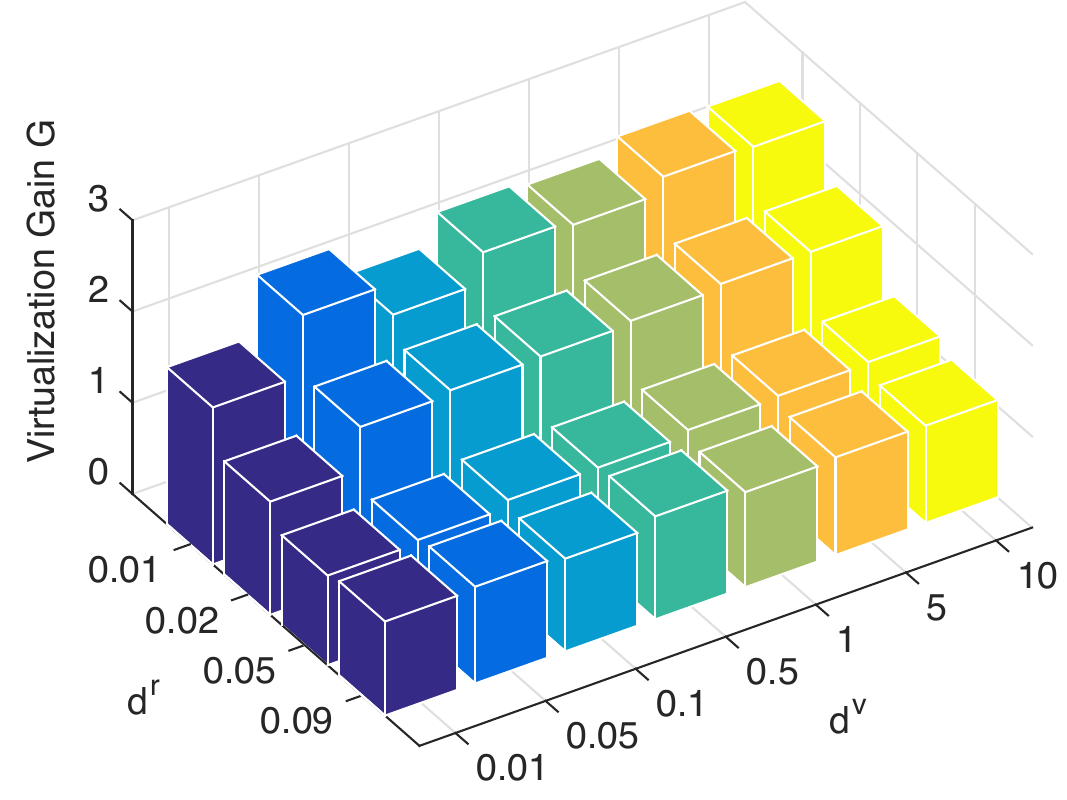}} \\
\caption{Virtualization gain $\mathcal{G}(d^{\msf{r}}, d^{\msf{v}}, k)$ as a function of $d^{\msf{r}}$  and $d^{\msf{v}}$, for $k=k_{\msf{opt}}$, in SPinV (a) vs. TWIST (b).}
\label{gains_bar}
\end{figure}

In analogy with Figures \ref{dr_subplot_SPinV} and \ref{dr_subplot_TWIST} presented in Section \ref{results_dr}, Figures \ref{dv_subplot_1_SPinV} and \ref{dv_subplot_1_TWIST} show the average positioning error $\bar{\epsilon}(d^{\msf{r}}_{\msf{min}}, d^{\msf{v}}, k)$ as a function of $k$ (Figures \ref{dv_subplot_1_SPinV_a} and \ref{dv_subplot_1_TWIST_a}) and the boxplot of the positioning error $\epsilon(d^{\msf{r}}_{\msf{min}}, d^{\msf{v}}, k_{\msf{opt}})$ (Figures \ref{dv_subplot_1_SPinV_b} and \ref{dv_subplot_1_TWIST_b}) in SPinV and TWIST testbeds, respectively.
\noindent Results show that, when $d^{\msf{r}}_{\msf{min}}$ is used, the introduction of virtual RPs significantly reduces the positioning error with respect to case of $d^{\msf{v}} = 0$. Results are generalized in Figure \ref{gains_bar}, showing the virtualization gain $\mathcal{G}(d^{\msf{r}}, d^{\msf{v}}, k_{\msf{opt}})$, as defined in Equation (\ref{gain_def}), for both testbeds, and confirm that the introduction of virtual RPs significantly improves positioning accuracy for low $d^{\msf{r}}$ as  $d^{\msf{v}}$ increases, while the advantage of introducing virtual RPs becomes negligible when a relatively large number of real RPs is available (high $d^{\msf{r}}$).
\begin{figure}
\centering
\includegraphics[trim=0 0 0 0, width=3in]{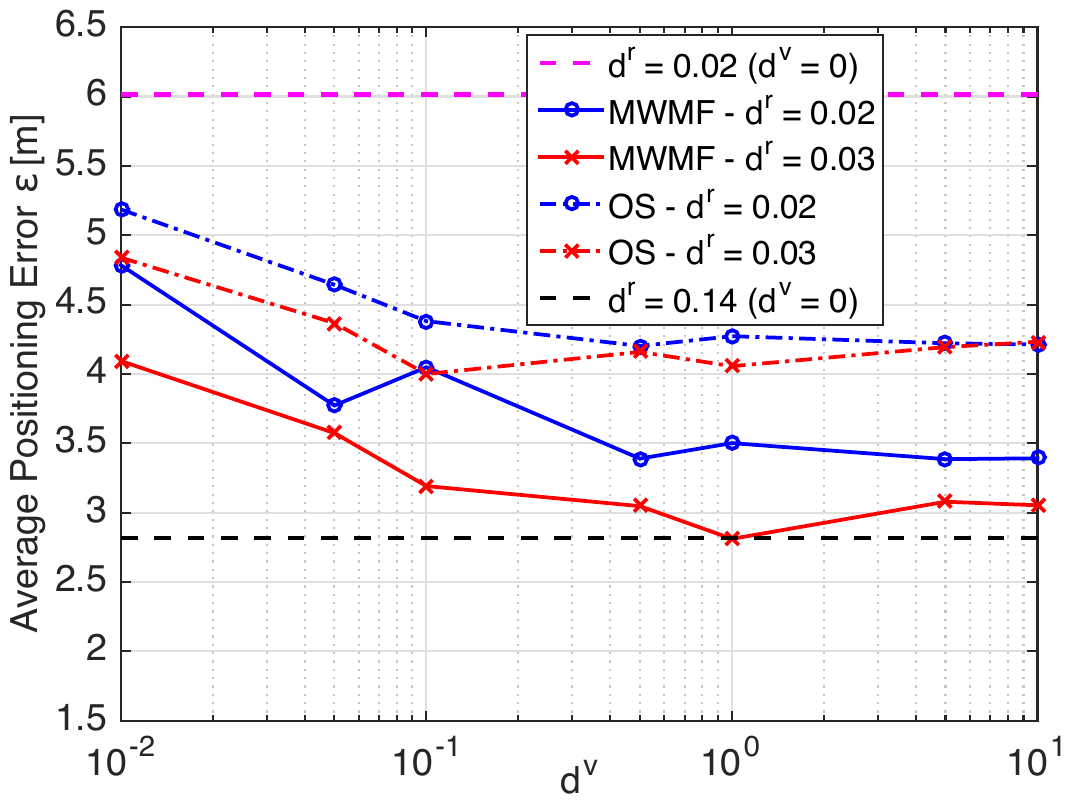}
\caption{Average positioning error $\bar{\epsilon}(d^{\msf{r}}, d^{\msf{v}}, k_{\msf{opt}})$ as a function of $d^{\msf{v}}$ for MWMF vs. OS models in SPinV, for $d^{\msf{r}}=0.02$ and $d^{\msf{r}}=0.03$; upper and lower bounds on positioning error observed for a real fingerprinting system with $d^{\msf{r}}=d^{\msf{r}}_{\msf{min}}$ vs. $d^{\msf{r}}=d^{\msf{r}}_{\msf{max}}$ are also shown.}
\label{Avg_Err_opt_SPinV}
\end{figure}
\begin{figure}
\centering
\includegraphics[trim=0 0 0 0, width=3in]{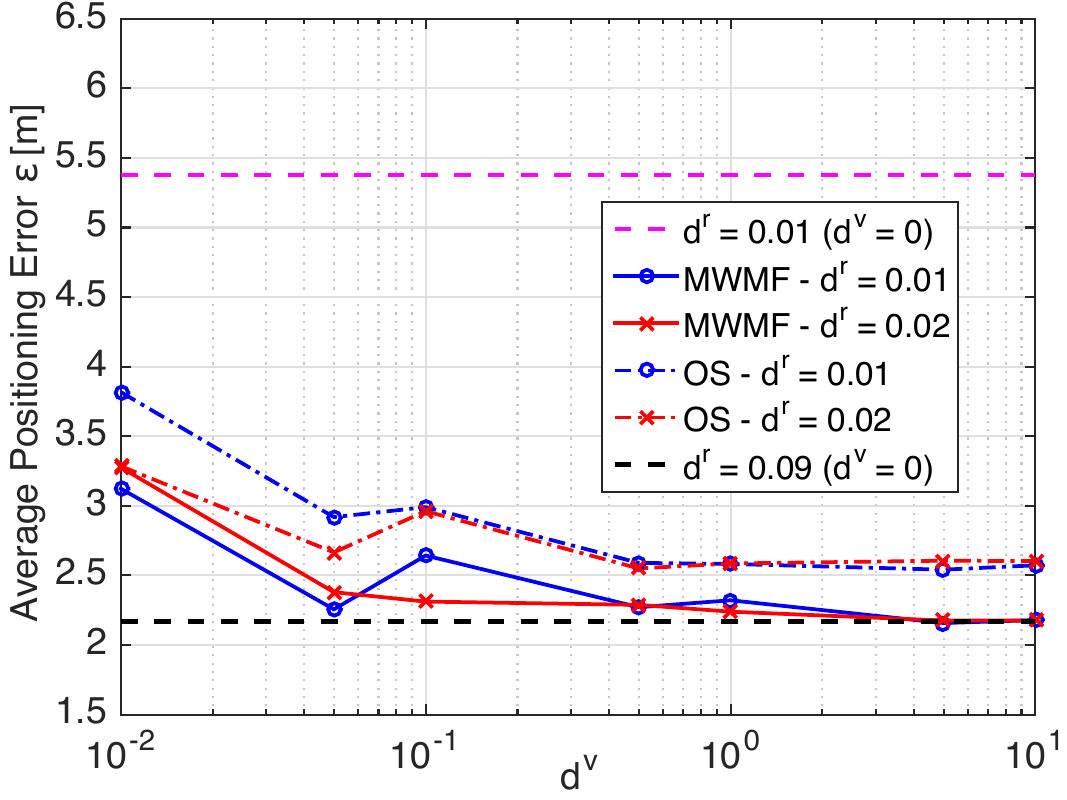}
\caption{Average positioning error $\bar{\epsilon}(d^{\msf{r}}, d^{\msf{v}}, k_{\msf{opt}})$ as a function of $d^{\msf{v}}$ for MWMF vs. OS models in TWIST, for $d^{\msf{r}}=0.01$ and $d^{\msf{r}}=0.02$; upper and lower bounds on positioning error observed for a real fingerprinting system with $d^{\msf{r}}=d^{\msf{r}}_{\msf{min}}$ vs. $d^{\msf{r}}=d^{\msf{r}}_{\msf{max}}$ are also shown.}
\label{Avg_Err_opt_TWIST}
\end{figure}

\noindent Low $d^{\msf{r}}$ values were thus selected for further investigation, with results presented in Figure \ref{Avg_Err_opt_SPinV} and Figure \ref{Avg_Err_opt_TWIST}, showing the average positioning error $\bar{\epsilon}(d^{\msf{r}}, d^{\msf{v}}, k_{\msf{opt}})$ as a function of  $d^{\msf{v}}$ for the two lowest values of $d^{\msf{r}}$ in SPinV vs. TWIST. Results are presented for both the MWMF model used in ViFi  and the OS model of \cite{Widyawan:2007}, \cite{Pulkkinen:2015}, as well as for a real fingerprinting system, with $d^{\msf{r}}=d^{\msf{r}}_{\msf{min}}$ and  $d^{\msf{r}}=d^{\msf{r}}_{\msf{max}}$ as upper and lower bounds for positioning error, respectively. Results indicate that in both testbeds the minimum average positioning error is obtained for $d^{\msf{v}}=d^{\msf{v}}_{\msf{max}}$, with the MWMF model adopted in ViFi leading to consistently better accuracy with respect to the OS model. Interestingly, results also show that the error measured with the MWMF model converges in all cases to $\bar{\epsilon}(d^{\msf{r}}_{\msf{max}}, 0, k_{\msf{opt}})$ (with the only exception of $\bar{\epsilon}(d^{\msf{r}}_{\msf{min}}, d^{\msf{v}}_{\msf{max}}, k_{\msf{opt}})$ in the SPinV testbed, about $60$ $\msf{cm}$ above $\bar{\epsilon}(d^{\msf{r}}_{\msf{max}}, 0, k_{\msf{opt}})$).\\
The results lead therefore to the conclusions that 1) the introduction of virtual fingerprints in ViFi effectively compensates the reduction of real ones with negligible effects on positioning accuracy, allowing to significantly reduce time required for the offline phase, and 2) the adoption of the MWMF model is instrumental in achieving this result.
\subsubsection{Empirical Derivation of $k_{\msf{est}}$}
\label{results_kest}
\begin{figure}
\centering
\subfloat[][\label{If}]
{\includegraphics[trim=0 0 0 0, width=1.68in]{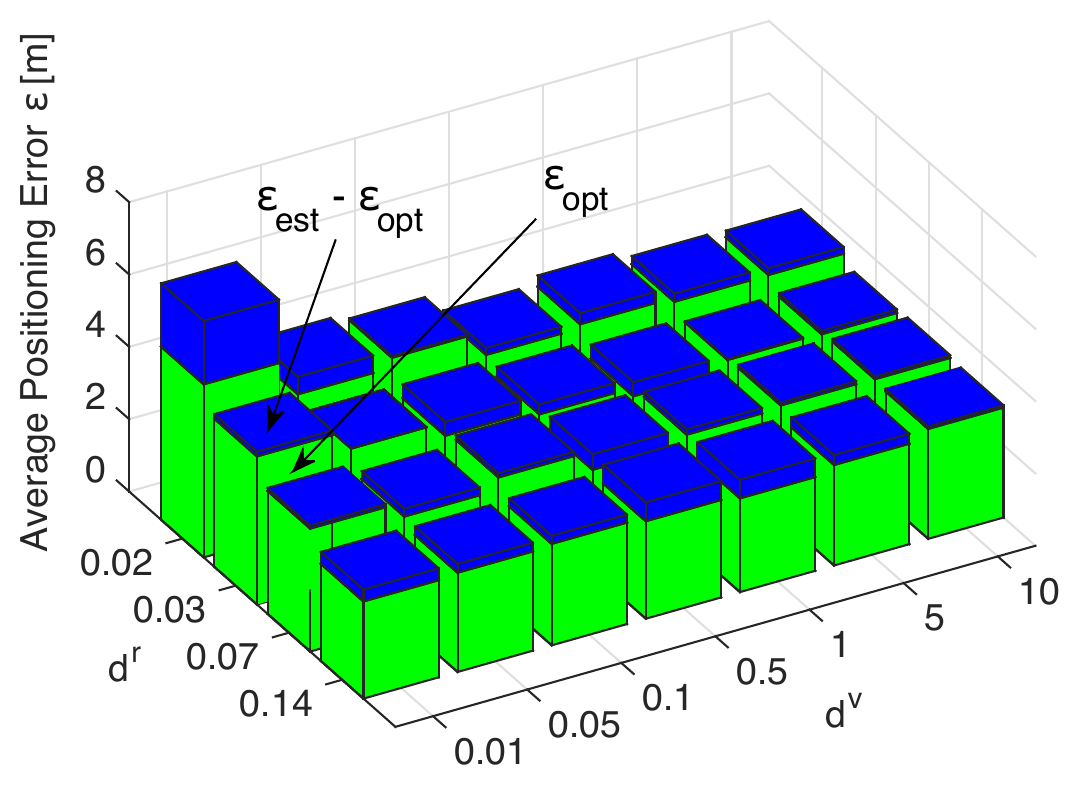}} \quad
\subfloat[][\label{IIf}]
{\includegraphics[trim=0 0 0 0, width=1.68in]{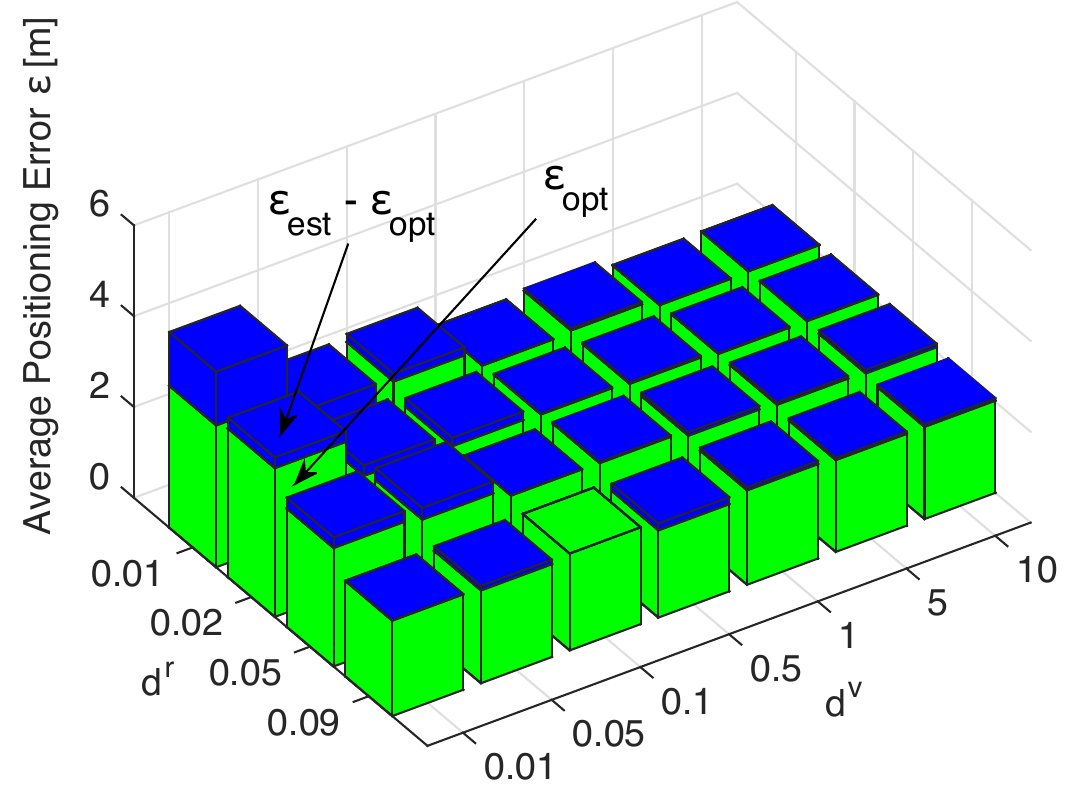}} \\
\caption{Comparison of average positioning errors $\bar{\epsilon}(d^{\msf{r}}, d^{\msf{v}}, k_{\msf{opt}})$ vs. $\bar{\epsilon}(d^{\msf{r}}, d^{\msf{v}}, k_{\msf{est}})$ as a function of $d^{\msf{r}}$ and $d^{\msf{v}}$, for $\alpha=0.05$, in SPinV (a) vs. TWIST (b).}
\label{epsilon_bar}
\end{figure}

The value of $\alpha$ guaranteeing a reliable estimation of $k_{\msf{opt}}$ was determined, for both environments, through the application of the procedure introduced in Section \ref{framework_kest}. Based on the analysis of the impact of $d^{\msf{v}}$ on the positioning error of Section \ref{results_dv}, the procedure was applied for $d^{\msf{v}}=d^{\msf{v}}_{\msf{max}}$.\\
\begin{figure}
\centering
\subfloat[][\label{Ie}]
{\includegraphics[trim=1cm 0 0 0, width=1.68in]{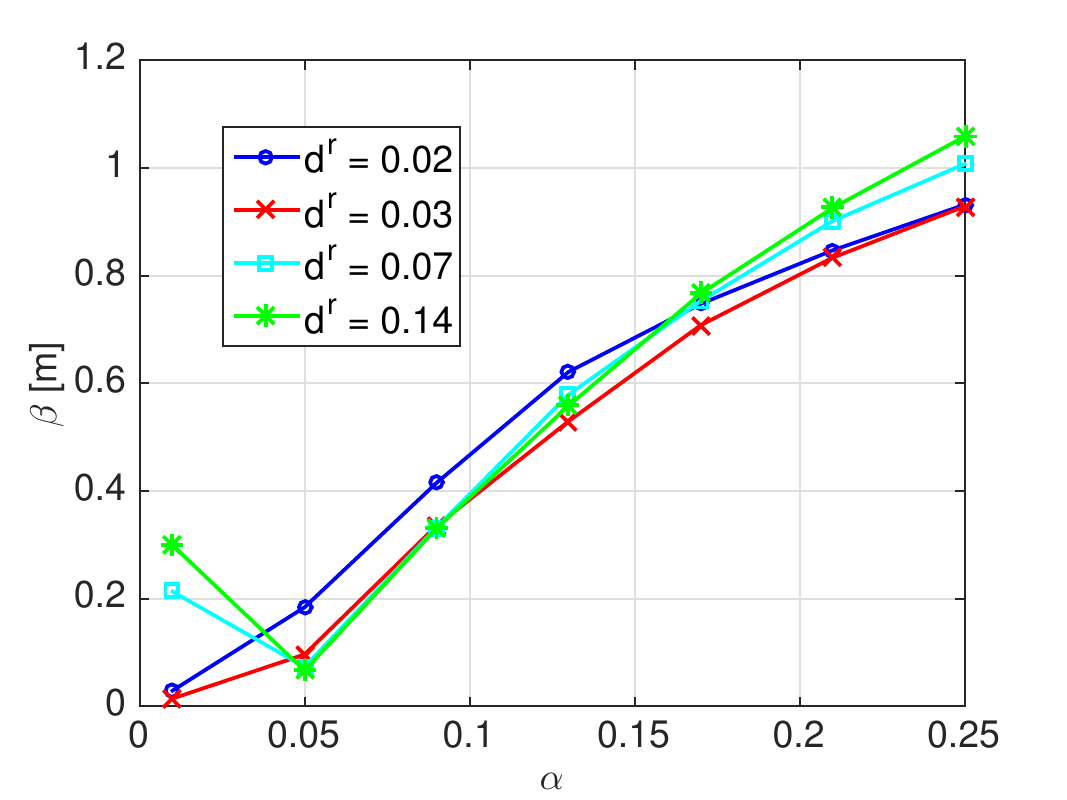}} \quad
\subfloat[][\label{IIe}]
{\includegraphics[trim=1cm 0 0 0, width=1.68in]{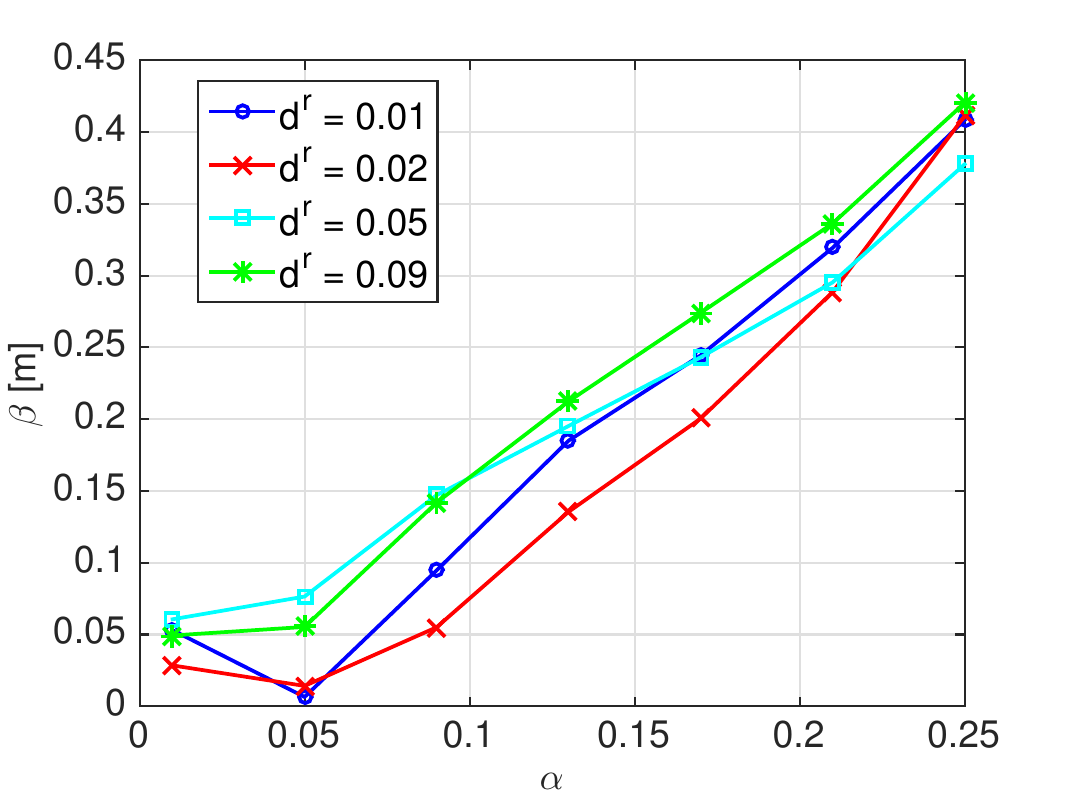}} \\
\caption{$\beta(d^{\msf{r}}, d^{\msf{v}}_{\msf{max}}, \alpha)$ as a function of $\alpha$, for $d^{\msf{r}}$ as in TABLE \ref{exp_settings} in SPinV (a) vs. TWIST (b).}
\label{coefficient}
\end{figure}Figure \ref{coefficient} presents $\beta(d^{\msf{r}}, d^{\msf{v}}_{\msf{max}}, \alpha)$ as a function of $\alpha$ for $d^{\msf{r}}$ as in TABLE \ref{exp_settings}.
Results show that $\alpha$ between $0.01$ and $0.05$ leads to values of $\beta(d^{\msf{r}}, d^{\msf{v}}_{\msf{max}}, \alpha)$ consistently close to its minimum, for all $d^{\msf{r}}$ values and in both SPinV and TWIST testbeds; this suggests that $k_{\msf{est}}=0.05\left(N^{\msf{r}} + N^{\msf{v}}\right)$  is a suitable approximation of $k_{\msf{opt}}$. The finding is supported by results shown in Figure \ref{epsilon_bar}, comparing the average positioning error $\bar{\epsilon}(d^{\msf{r}}, d^{\msf{v}}, k_{\msf{est}})$ vs. $\bar{\epsilon}(d^{\msf{r}}, d^{\msf{v}}, k_{\msf{opt}})$ in the SPinV vs. TWIST testbeds. The small difference between the average positioning errors corroborates the reliability of the proposed $k_{\msf{opt}}$ estimator and its applicability, with $\alpha$ set to $0.05$, to different environments.


\subsection{Crowdsourcing-like Scenario}
\label{results_crowdsourcing}
\begin{figure}[!h]
\centering
\includegraphics[trim=0 0 0 0, width=3in]{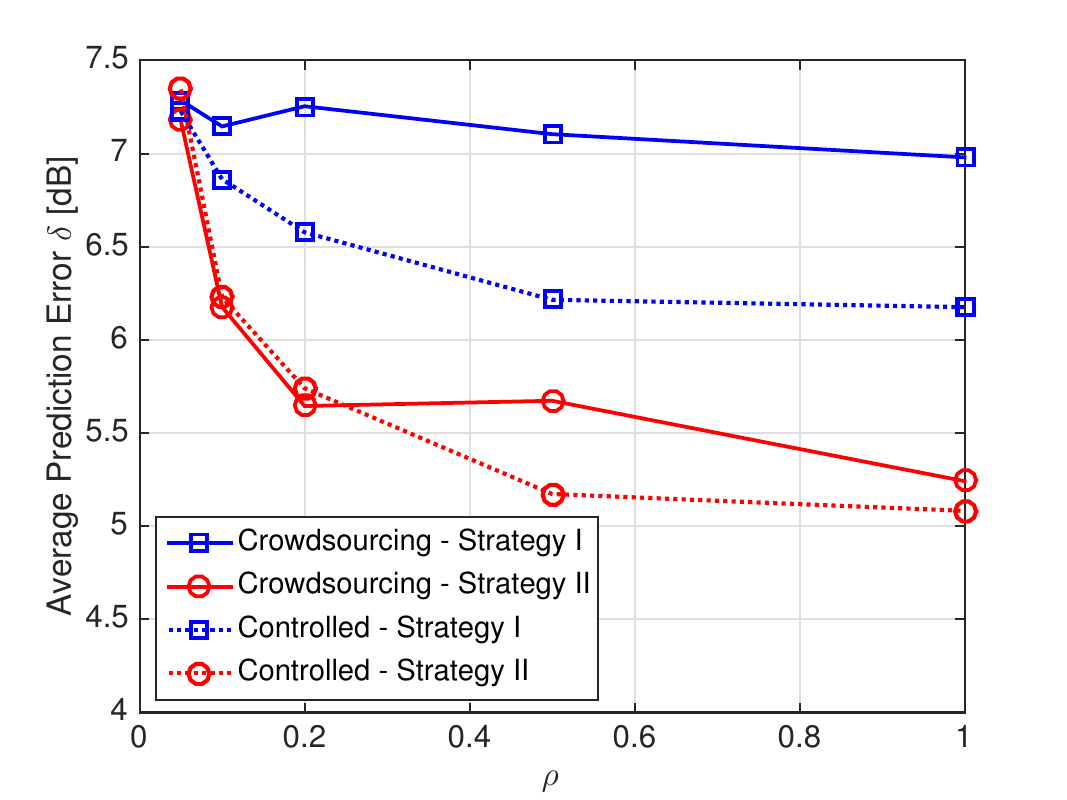}
 \caption{Average prediction error $\bar{\delta}(\rho)$ as a function of $\rho$ for selection Strategy I vs. II in the Crowdsourcing-like scenario.}
\label{delta_m_cr}
\end{figure}

Figure \ref{delta_m_cr} reports the MWMF average prediction error $\bar{\delta}(\rho)$ for the SPinV testbed in the Crowdsourcing-like scenario, in comparison with the error obtained in the Controlled scenario, already presented in Figure \ref{delta_m_average_error}. Figure highlights that the use of less stable measurements, inherent to crowdsourcing, leads to a degradation in prediction accuracy.

\begin{figure}[!h]
\centering
\includegraphics[trim=0 0 0 0, width=3in]{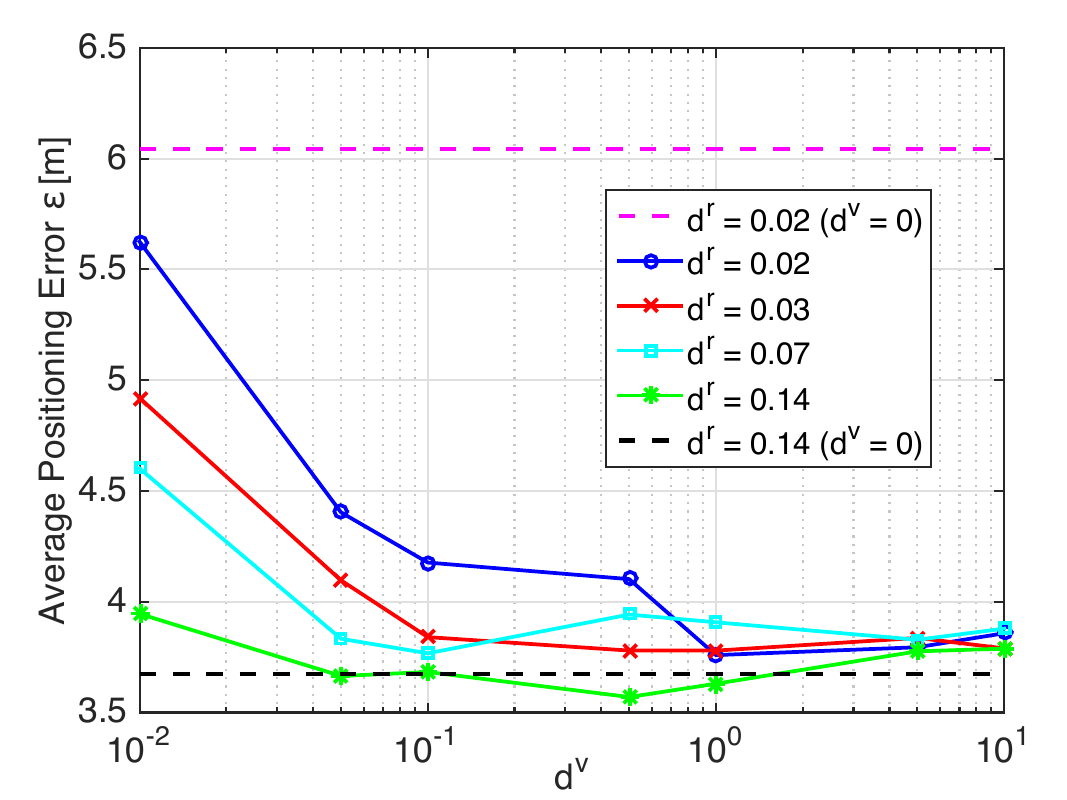}
 \caption{Average positioning error $\bar{\epsilon}(d^{\msf{r}}, d^{\msf{v}}, k_{\msf{opt}})$ as a function of $d^{\msf{v}}$, for $d^{\msf{r}}$ as in TABLE \ref{exp_settings}, in the Crowdsourcing-like scenario; upper and lower bounds on positioning error observed for a real fingerprinting system with $d^{\msf{r}}=d^{\msf{r}}_{\msf{min}}$ vs. $d^{\msf{r}}=d^{\msf{r}}_{\msf{max}}$ are also shown.}
\label{epsilon_est_cr}
\end{figure}

\begin{figure}
\centering
\subfloat[][\label{G_cr}]
{\includegraphics[trim=0 0 0 0, width=1.68in]{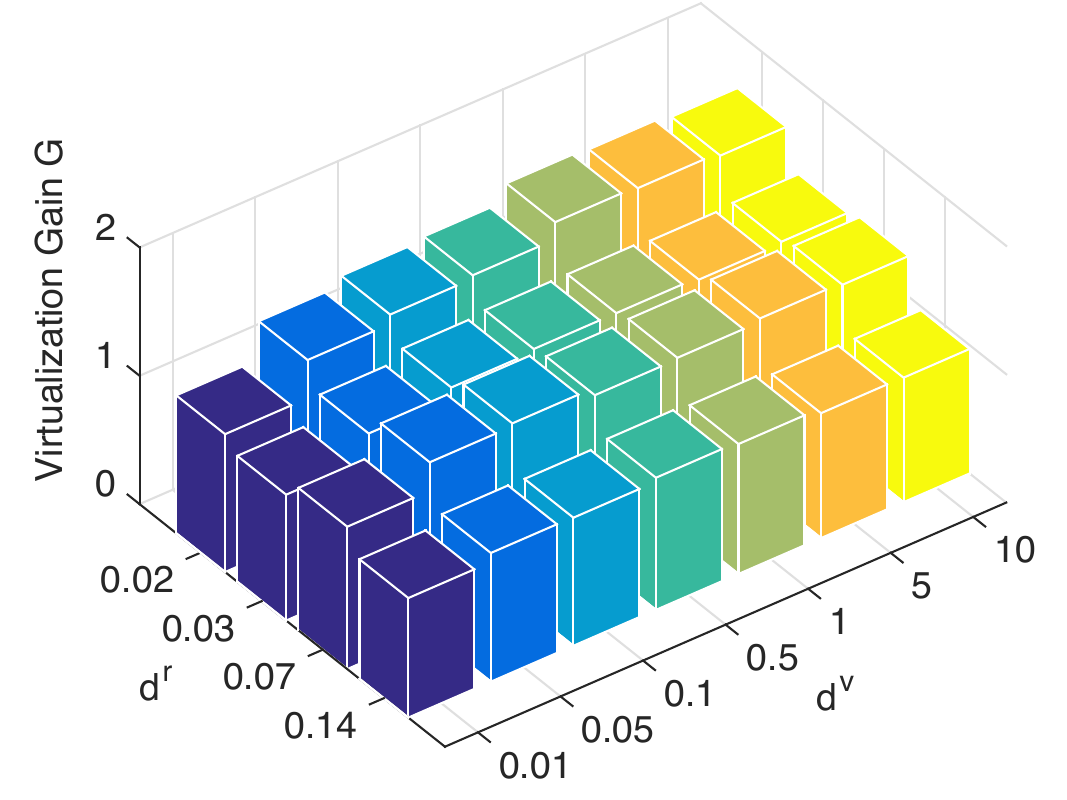}} \quad
\subfloat[][ \label{epsilon_bar_cr}]
{\includegraphics[trim=0 0 0 0, width=1.68in]{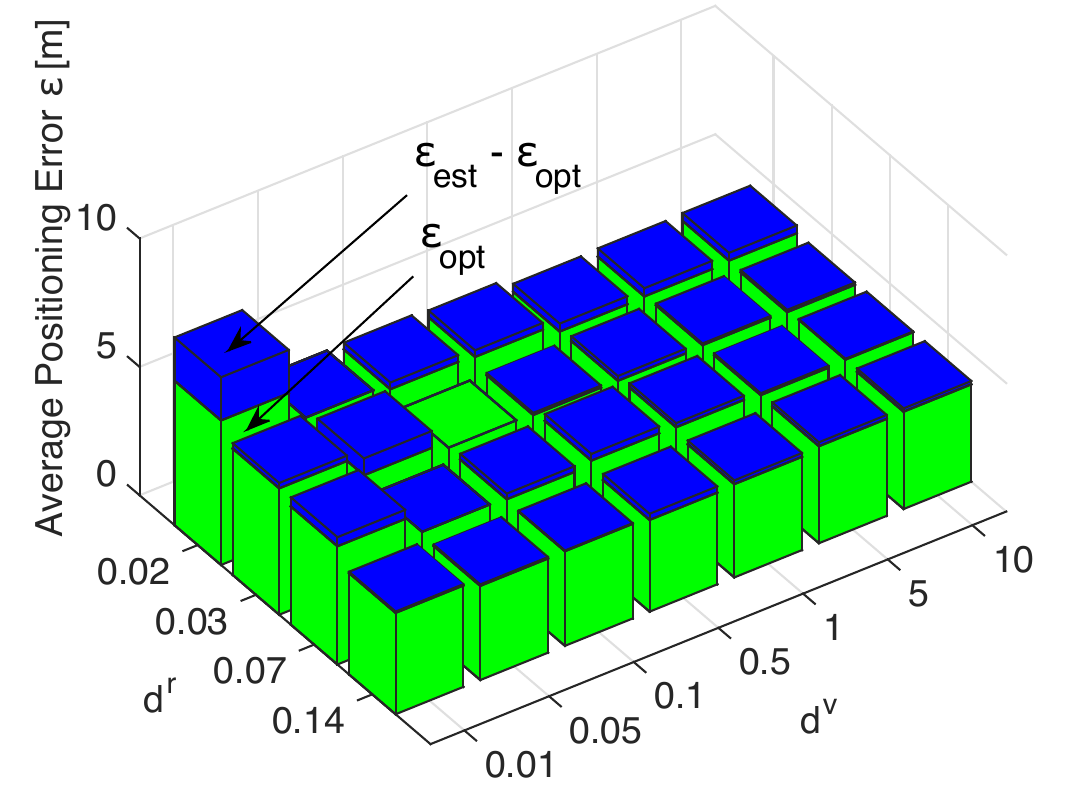}} \\
\caption{Virtualization gain $\mathcal{G}(d^{\msf{r}}, d^{\msf{v}}, k)$ as a function of $d^{\msf{r}}$ and $d^{\msf{v}}$, for $k=k_{\msf{opt}}$ (a), and comparison of average positioning error $\bar{\epsilon}(d^{\msf{r}}, d^{\msf{v}}, k_{\msf{opt}})$ and $\bar{\epsilon}(d^{\msf{r}}, d^{\msf{v}}, k_{\msf{est}})$ for $\alpha=0.05$ (b), as a function of $d^{\msf{r}}$, $d^{\msf{v}}$ in the Crowdsourcing-like scenario.}
\label{cr}
\end{figure}

\noindent The same holds true for the average positioning error $\bar{\epsilon}(d^{\msf{r}}, d^{\msf{v}}, k_{\msf{opt}})$ presented in Figure \ref{epsilon_est_cr}. Comparison against Figure \ref{Avg_Err_opt_SPinV} confirms that the use of crowdsourced measurements decreases the positioning accuracy, with an increase of the error lower bound $\bar{\epsilon}(d^{\msf{r}}_{\msf{max}}, 0, k_{\msf{opt}})$ of about $90$ $\msf{cm}$. ViFi, however, still reaches the new lower bound, confirming its capability to compensate the lack of an exhaustive offline measurement phase with the introduction of virtual fingerprints. The virtualization gain $\mathcal{G}(d^{\msf{r}}, d^{\msf{v}}, k_{\msf{opt}})$, reported in Figure \ref{G_cr}, shows similar trends with respect to the Controlled scenario seen in Figure \ref{gains_bar}, with higher virtualization gains for low $d^{\msf{r}}$, high $d^{\msf{v}}$ values.

\noindent Finally, Figure \ref{epsilon_bar_cr} compares, in analogy with Figure \ref{epsilon_bar}, the average positioning errors $\bar{\epsilon}(d^{\msf{r}}, d^{\msf{v}}, k_{\msf{opt}})$ and $\bar{\epsilon}(d^{\msf{r}}, d^{\msf{v}}, k_{\msf{est}})$, with $k_{\msf{est}}$ evaluated with $\alpha=0.05$: results confirm that the selection of $5 \%$ of the total amount of RPs as value of $k$ leads to an average positioning error within $5\%$ of the positioning error obtained for $k_{\msf{opt}}$. In light of the above results, it is possible to conclude that ViFi is applicable to both controlled and crowdsourcing scenarios, and that crowdsourcing might prove a valuable option to further reduce the offline phase. Given the significant losses in terms of positioning accuracy caused by the introduction of unreliable measurements, further studies are however required to determine the full extent of the impact of crowdsourced data on virtual fingerprinting systems.
\subsection{Summary of key experimental results}
\label{summary_results}
Key results of the analysis presented in the previous subsections can be summarized as follows:
\begin{itemize}
\item positioning error decreases as $d^{\msf{v}}$ increases, but values of $d^{\msf{v}} > 10$ RPs$/\msf{m}^2$ do not provide significant additional improvement (Figures \ref{dv_subplot_1_SPinV} and \ref{dv_subplot_1_TWIST});
\item positioning error decreases as $d^{\msf{r}}$ increases, but values of $d^{\msf{r}} > 0.03$ RPs$/\msf{m}^2$  do not provide additional improvement \emph{if virtual fingerprinting is used} (Figures \ref{Avg_Err_opt_SPinV} and  \ref{Avg_Err_opt_TWIST});
\item $k_{\msf{opt}}$ is directly proportional to $d^{\msf{r}}+ d^{\msf{v}}$, supporting the validity of Equation (\ref{k}) (Figures \ref{dv_subplot_1_SPinV_b} and \ref{dv_subplot_1_TWIST_b}), and $\alpha=0.05$  consistently leads to a good approximation of $k_{\msf{opt}}$ across different values of $d^{\msf{r}}$ and $d^{\msf{v}}$ (Figure \ref{epsilon_bar}).
\end{itemize}
Such results led to the implementation guidelines detailed in Section \ref{guidelines}.
\section{Implementation Guidelines}
\label{guidelines}
The consistency in the ViFi behavior across two different environments suggests the following set of implementation guidelines in a generic environment of area $|\mathcal{A}|$:
\begin{enumerate}
\item A density of real RPs of $d^{\msf{r}}=0.03$ RPs$/\msf{m}^2$ is sufficient to obtain a reliable RSS prediction using the MWMF model, and the corresponding generation of virtual RPs. Repeated measurements for each RP should be collected and averaged, in order to mitigate the impact of unstable measurements on positioning error.
\item The average RSS prediction error $\bar{\delta}$ is a reliable indicator for choosing the selection strategy leading to the minimization of the average positioning error $\bar{\epsilon}$.
\item A density of virtual RPs from $1$ to $10$ RPs$/\msf{m}^2$ is sufficient to achieve a positioning accuracy comparable with a real fingerprinting system operating in the same environment with $d^{\msf{r}}=d^{\msf{r}}_{\msf{max}}$.
\item The selection of $k_{\msf{est}}=\lceil0.05(d^{\msf{r}} + d^{\msf{v}})|\mathcal{A}|\rceil$ in the W$k$NN estimator guarantees an average positioning error within $5\%$ of the error achieved with $k=k_{\msf{opt}}$.          
\end{enumerate}
The above guidelines were heuristically derived, and their validity cannot be proved for all testbeds and all environments; however, the following observations support the claim of applicability across different testbeds:
\begin{enumerate}
\item SPinV and TWIST do not share any hardware or software component, barring the possibility that the results are depending on a specific combination of infrastructure, devices, or software used during data collection.
\item SPinV and TWIST are deployed in different environments, with different topological characteristics, and different number and positions of APs. These differences are clearly reflected by the different accuracy achievable in the two testbeds, with TWIST consistently leading to better accuracy compared to SPinV: nonetheless, the above guidelines, when applied to the two testbeds, lead to an average positioning error very close (within $5.5\%$) to the achievable minimum.
\item As mentioned above, the guidelines were applied to a third testbed, w-iLab.t I, and experimental results fully corroborate the proposed guidelines \cite{ViFi:supplemental}.  
\end{enumerate}
\section{ViFi vs. traditional fingerprinting}
\label{requirements}
\begin{description}[style=unboxed, leftmargin = 0cm]
\item[Information requirements] -- ViFi requires two additional information pieces when compared to traditional fingerprinting: 1) the position of the APs and 2) the transmit power of the APs. As of 1), the adoption of Simultaneous Location And Mapping (SLAM) techniques might be considered in scenarios where the position of the APs is not known in advance, with a possible loss in accuracy \cite{BaiDurrant:2006}. Regarding 2), the transmit power may be included in the set $\{\mathcal{S}\}$ in the optimization procedure defined by Equation (\ref{fitting}), but Strategy II is preferable in this case, unless one assumes that $W^{\msf{EIRP}}_{\msf{TX}},$ although unknown, is the same for all APs.
\item[Positioning accuracy] -- The positioning accuracy in WiFi fingerprinting depends on the characteristics of the environment, as shown in \cite{Lemic:2016},\cite{VanHaute:2016}. Although an average positioning error of about $1$ $\msf{m}$ was achieved in an indoor area of roughly $10 \times 10$ $\msf{m}^2$ \cite{Lym:2015}, positioning error increases as the environment size increases: in large and complex indoor environments, such as those considered in this work, an average positioning error in the order of $2$ $\msf{m}$ is in line with current state of the art. Indeed, similar positioning errors have been recently demonstrated for several localization solutions in TWIST \cite{Lemic:2015} and other environments \cite{Kumar:2016}.
\item[Complexity and Scalability] -- The complexity of the offline phase in ViFi is determined by the least square fitting procedure (Equation (\ref{fitting})), and by the generation of virtual RPs (Equations (\ref{mwmf_gen})-(\ref{cost213})). Following \cite{Simon:2002}, the single fitting procedure of Strategy I has complexity $O(|\mathcal{S}|(N^{\msf{r}}L)^{2})$, while the $L$ different fitting procedures of Strategy II have complexity $O(|\mathcal{S}|(N^{\msf{r}})^{2})$ each. As for the generation of virtual RPs, given the total number $N^{\msf{v}}L$ of RSS values to be predicted, and denoted with $N^{\msf{MAX}}_{\msf{2D}}$ and $N^{\msf{MAX}}_{\msf{f}}$ the maximum number of 2D objects and floors obstructing the $N^{\msf{v}}L$ links, the complexity of the virtual RPs generation is $O(N^{\msf{MAX}}_{\msf{2D}}N^{\msf{v}}L)$ in terms of multiplications (assuming $N^{\msf{MAX}}_{\msf{2D}} > N^{\msf{MAX}}_{\msf{f}}$). The complexity of the ViFi offline phase is, by definition, larger than in traditional fingerprinting, where no computations are required to fill the database. This increase in complexity is however largely compensated by the dramatic reduction in time and efforts required for RSS collection. As a reference, the offline phase of ViFi in the SPinV testbed would require about $45$ minutes instead of about $7$ hours. As a conclusion, ViFi outperforms traditional fingerprinting in terms of system implementation scalability. A better scalability enables an efficient deployment of ViFi in scenarios where traditional fingerprinting would be unpractical due to the size of the area or to the need for frequent radiomap updates to cope with a variable environment.\\
Regarding the online phase, according to the analysis presented in \cite{Zuo:2008}, the complexity of the W$k$NN algorithm is determined by a) the computation of the Euclidean distances between the target RSS fingerprint and the RPs ($NL$ multiplications), b) the selection of the $k$ nearest neighbors, ($Nk$ comparisons), and c) the application of the weights to the $k$ neighbors ($k$ multiplications). Since in this work a linear dependency on $N$ is proposed for $k$, the online complexity is $O(NL)$ multiplications and $O(N^2)$ comparisons. The complexity of the ViFi online phase is exactly the same as of traditional fingerprinting adopting a W$k$NN algorithm, but ViFi may require a slightly longer execution time due to the larger set of real $+$ virtual RPs. However, several optimization mechanisms can be easily applied to ViFi, aiming to improve its usage scalability. For example, in so-called \emph{two-step} W$k$NN schemes \cite{Caso:Sens:2015}, the complexity can be significantly reduced by performing a clustering of RPs via Affinity Propagation algorithm, and adopting \emph{coarse} and \emph{fine} positioning steps for the online phase. 
\item[Vulnerability] -- A plethora of works investigate the vulnerabilities of fingerprinting systems \cite{He:2016}. Two vulnerabilities particularly relevant to ViFi can be identified: environment variations over time and device heterogeneity. \emph{Environmental changes over time} may impact positioning accuracy by invalidating the RSS data, requiring thus a radiomap update. In this respect, ViFi provides two major improvements. First, if the environment change is restricted to a specific topological variation (e.g. a single wall being removed), the underlying MWMF model allows to update the ViFi radiomap by simply regenerating the set of virtual RPs using the new topology, without the need for a new channel training. Second, even if major changes occur, calling for a new channel training, ViFi requires a significantly lower number of real RPs to operate, thus reducing the time and effort required to collect data and update the radiomap, so to restore the system to its best performance. \emph{Measurement mismatch due to device heterogeneity} may also negatively impact accuracy. Solutions proposed to address this issue in real fingerprinting systems, such as using the difference between RSS values so to erase biases introduced by specific hardware \cite{Hossain:2013}, or removing from each RSS value the average RSS received from all APs at the same location \cite{Wang:2013} might be easily applied to ViFi as well. Moreover, the low number of measurements to be collected in ViFi enables a new solution, unpractical in traditional fingerprinting: multiple sets of measurements with different devices can be collected during the offline phase while still saving significant time and effort with respect to the extensive data collection campaign of a real fingerprinting system. In the online phase, the database that best fits the target device can be then adopted to mitigate the impact of device heterogeneity.  
\end{description}

\section{Conclusion}
\label{conclusion}
In this work a virtual fingerprinting indoor positioning system, referred to as ViFi, has been proposed. ViFi uses the empirical MWMF indoor propagation model for the generation of virtual RPs and a deterministic Euclidean W$k$NN algorithm to infer the target location.\\
The performance of ViFi was experimentally evaluated in multiple independent testbeds, and compared with previous proposals in the literature. Results show that ViFi outperforms virtual fingerprinting systems using simpler propagation models, and provides the same accuracy of a real fingerprinting system while guaranteeing up to a sevenfold reduction in time and efforts for measurements.\\%
A set of guidelines for the selection of offline and online ViFi parameters was also proposed, that saves the additional efforts related to the testing phase typically required for tuning a WiFi fingerprinting system.\\ 
This work opens the way for further research on several topics, including: a) adoption of other online estimation algorithms in place of the W$k$NN algorithm, providing better accuracy and/or lower complexity, and verification of the guidelines in Section \ref{guidelines}; b) improved propagation modeling, taking into account device heterogeneity and orientation, particularly relevant in a crowdsourced scenario; c) derivation of enhanced models for the design and analysis of WiFi fingerprinting systems. 

\ifCLASSOPTIONcompsoc
\else
\fi


\ifCLASSOPTIONcaptionsoff
  \newpage
\fi



%

\vspace{-1.5cm}
\begin{IEEEbiography}[{\includegraphics[width=1in,height=1.25in,clip,keepaspectratio]{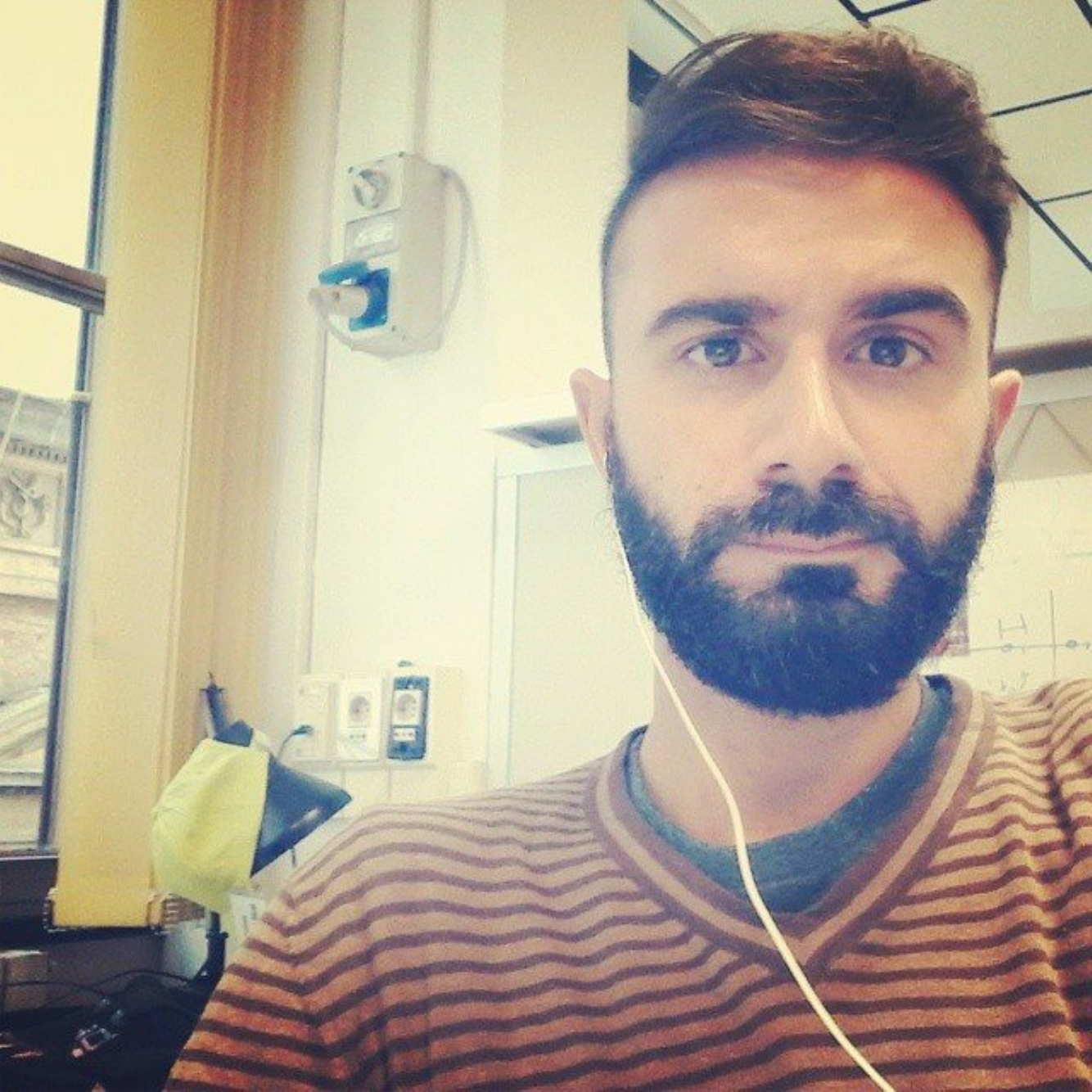}}]{Giuseppe Caso}
is a Postdoctoral Fellow at the MOSAIC Dept., SimulaMet. In 2016, he received the Ph.D. degree from Sapienza University of Rome, where he was a Postdoctoral Fellow until 2018. From 2012 to 2018, he has held visiting positions at Leibniz University of Hannover, King's College London, Technical University of Berlin, and Karlstad University. His research interests include cognitive, context-aware, and distributed communications, and WiFi/UWB positioning systems. He is an IEEE Member. 
\end{IEEEbiography}
\vspace{-1.5cm}
\begin{IEEEbiography}[{\includegraphics[width=1in,height=1.25in,clip,keepaspectratio]{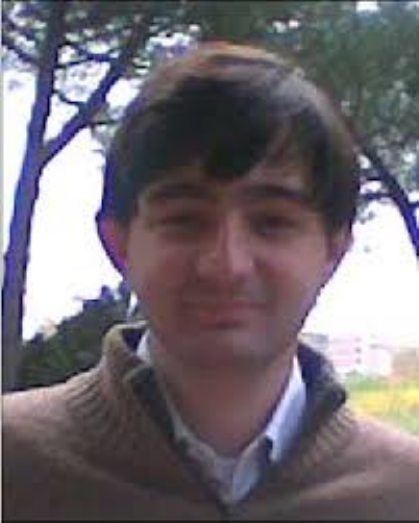}}]{Luca De Nardis}
(M'98) is an Assistant Professor with the DIET Dept., Sapienza University of Rome. He received the Ph.D. degree from Sapienza University of Rome in 2005. In 2007, he was a Postdoctoral Fellow with the EECS Dept., University of California, Berkeley. He authored or co-authored over 100 international peer-reviewed publications. His research interests focus on UWB and cognitive communications, medium access control, routing protocols and positioning systems for wireless networks.  
\end{IEEEbiography}
\vspace{-1.5cm}
\begin{IEEEbiography}[{\includegraphics[width=1in,height=1.25in,clip,keepaspectratio]{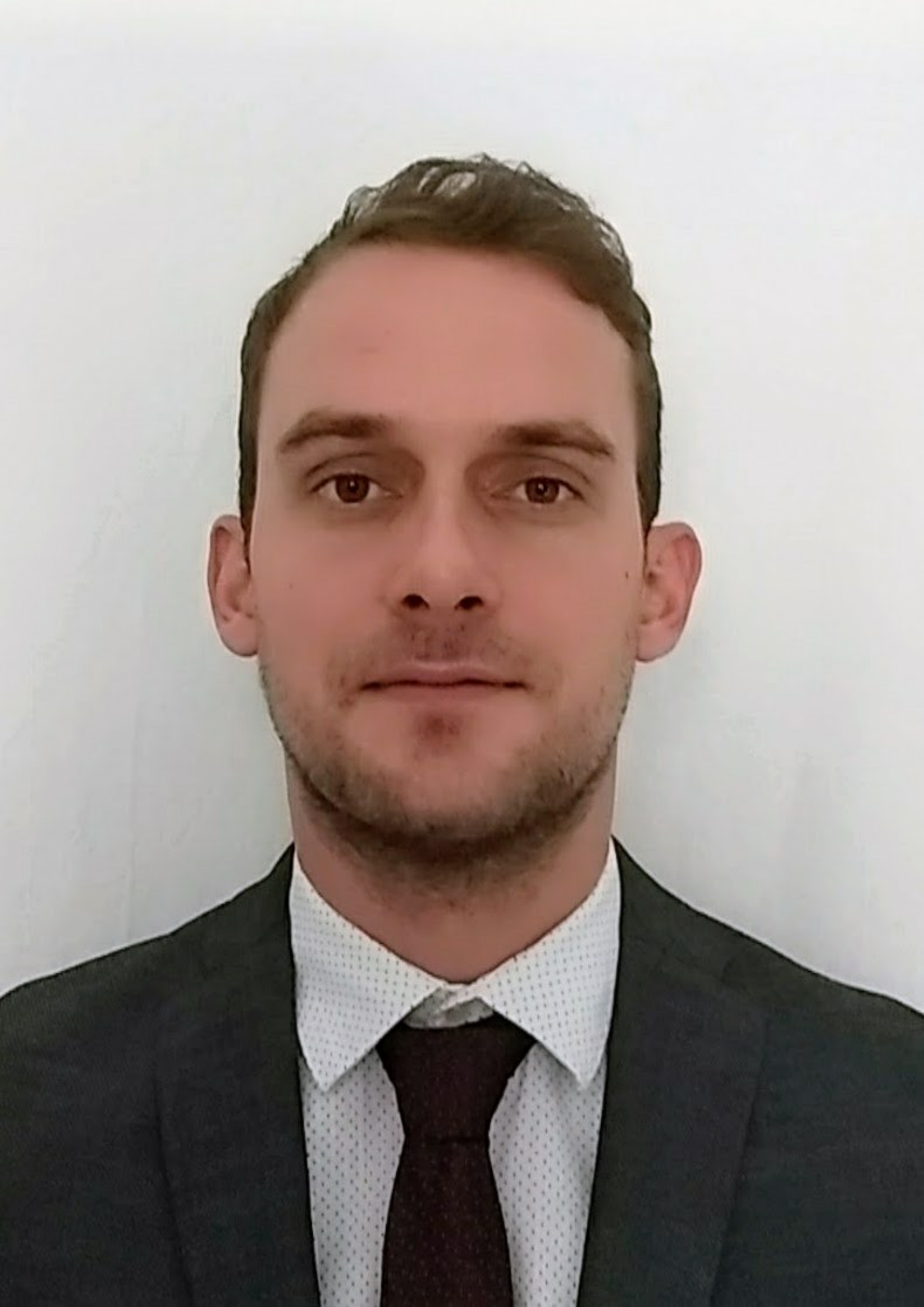}}]{Filip Lemic}
is a Postdoctoral researcher at the IDLab, University of Antwerp, and senior researcher at the imec institute. In 2017, he received his Ph.D. from Technical University of Berlin. He has held visiting positions at the University of California, Berkeley, in 2015 and 2016. He co-authored more than 30 international peer-reviewed publications. His research interests include context-aware, mmWave, and opportunistic communications, as well as testbeds for evaluation of wireless networks.
\end{IEEEbiography}
\vspace{-1.5cm}
\begin{IEEEbiography}[{\includegraphics[width=1in,height=1.25in,clip,keepaspectratio]{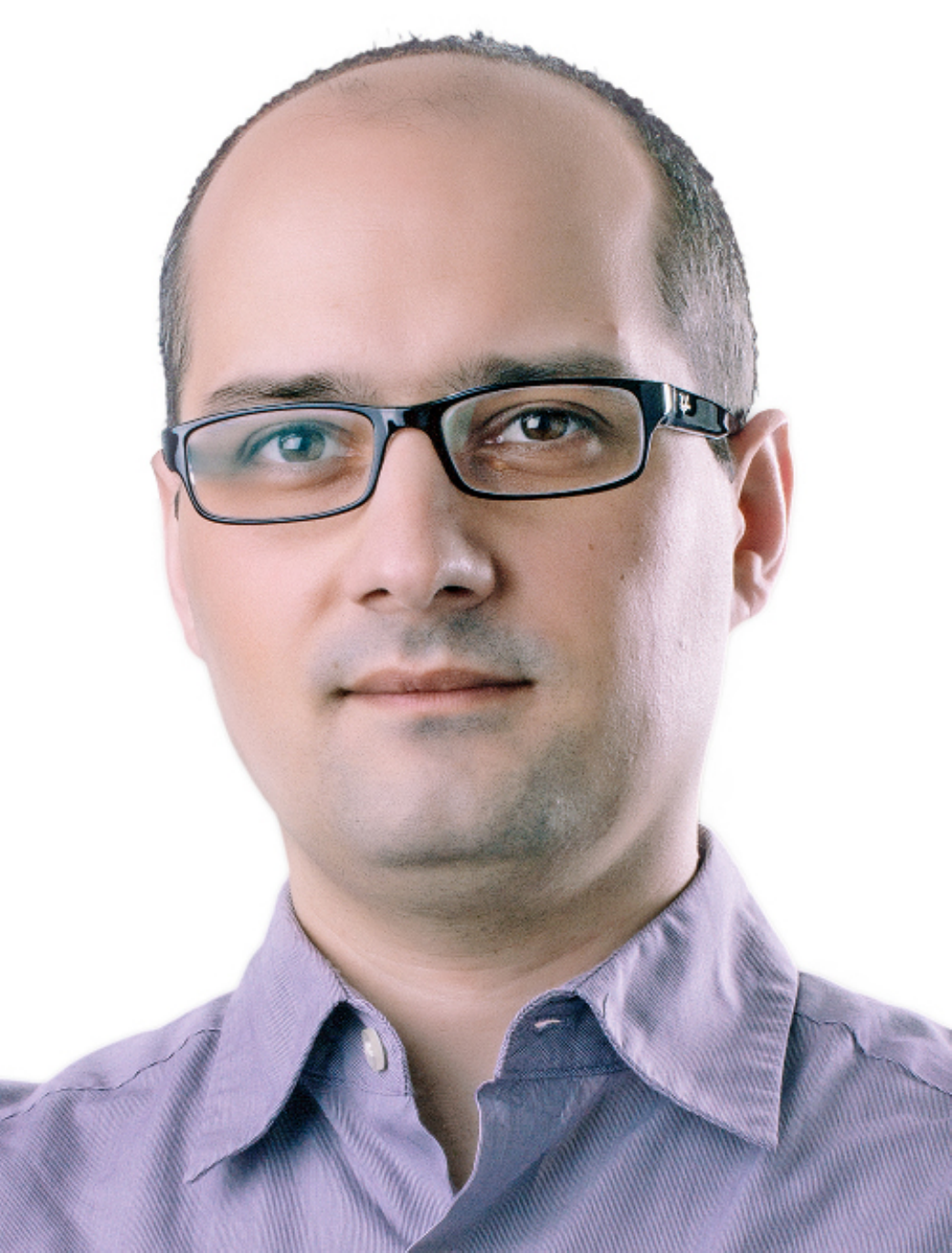}}]{Vlado Handziski}
received the M.Sc. degree from Ss. Cyril and Methodius University in Skopje, and Ph.D. degree from Technical University of Berlin, in 2002 and 2011. He is a Senior Researcher in the TKN Group, Technical University of Berlin, where he coordinates the activities in the areas of sensor networks, cyber-physical systems, and Internet of Things. He is also serving as Interim Professor at the chair for Embedded Systems at Technical University of Dresden. He is an IEEE Member.
\end{IEEEbiography}
\vspace{-1.5cm}
\begin{IEEEbiography}[{\includegraphics[width=1in,height=1.25in,clip,keepaspectratio]{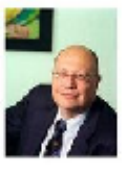}}]{Adam Wolisz}
received his degrees (Diploma 1972, Ph.D. 1976, Habil. 1983) from Silesian University of Technology, Gliwice. In 1993 he has been appointed Professor of EE\&CS at the Berlin University of Technology (TU Berlin) where he has founded and led for 25 years the TKN Group. Since 2005-2017 he has also been Adjunct Professor at EECS Dept., University of California, Berkeley. As by Autumn 2018 he has retired, and continues research activities at TU Berlin as well as Berkeley Wireless Research Center (Visiting scholar) while being active as Fellow within the Einstein Center Digital Future in Berlin. His research interests are in architectures and protocols of communication networks. He is an IEEE Senior Member. 
\end{IEEEbiography}
\vspace{-1.5cm}
\begin{IEEEbiography}[{\includegraphics[width=1in,height=1.25in,clip,keepaspectratio]{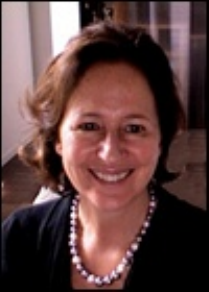}}]{Maria-Gabriella Di Benedetto} obtained her Ph.D. degree in 1987 from Sapienza University of Rome. In 1991, she joined the Faculty of Engineering of Sapienza University, where she is a Full Professor of telecommunications. She has held visiting positions at the Massachusetts Institute of Technology, the University of California, Berkeley, and the University of Paris XI. In 1994, she received the Mac Kay Professorship award from the University of California, Berkeley. Her research interests include wireless communication systems, in particular impulse radio communications, and speech. She is an IEEE Fellow.
\end{IEEEbiography}




\end{document}